\documentstyle[aps,multicol]{revtex}
\def\tr{\mathop{\rm Tr}\nolimits}
\def\det{\mathop{\rm Det}\nolimits}
\def\str{\mathop{\rm STr}\nolimits}
\def\sdet{\mathop{\rm SDet}\nolimits}
\def\ad{\mathop{\rm ad}\nolimits}
\def\Ad{\mathop{\rm Ad}\nolimits}
\def\half{{\textstyle{1\over 2}}}
\begin{document}
\draft
\title{
Single-particle Green's functions of the Calogero-Sutherland
model at couplings $\lambda$ = 1/2, 1 and 2}
\author{M. R. Zirnbauer}
\address{Institut f\"{u}r Theoretische Physik,
Universit\"{a}t zu K\"{o}ln, Z\"{u}lpicherstr. 77,
50937 K\"{o}ln, Germany}
\author{F. D. M. Haldane}
\address{Department of Physics, Princeton University,
Princeton, New Jersey 08544}
\date{April 26, 1995}
\maketitle
\begin{abstract}
At coupling strengths $\lambda$ = 1/2, 1, or 2, the
Calogero-Sutherland model (CSM) is related to Brownian motion in a
Wigner-Dyson random matrix ensemble with orthogonal, unitary, or
symplectic symmetry. Using this relation in conjunction with
superanalytic techniques developed in mesoscopic conductor physics, we
derive an exact integral representation for the CSM two-particle
Green's function in the thermodynamic limit. Simple closed expressions
for the single-particle Green's functions are extracted by separation
of points. For the advanced part, where a particle is added to the
ground state and later removed, a sum of two contributions is found:
the expected one with just one particle excitation present, plus an
extra term arising from fractionalization of the single particle into
a number of elementary particle and hole excitations.
\end{abstract}
\pacs{71.10.+x, 75.10.Am}
\begin{multicols}{2}
\section{Introduction}
While the study of integrable models of interacting particle systems
in one dimension has a long history, going back to Bethe\cite{bethe},
it has proved difficult to extract explicit expressions for their
correlation functions from the solutions. Recently, however, the
correlation functions of one class of such models, those with
inverse-square interactions, have started to become available.
These models have a special simplicity as compared to the
integrable models of the Bethe type, and combine some of the
special properties of an ideal gas with non-trivial statistics.

The original model of this type is the Calogero-Sutherland
model\cite{calogero,sutherland} (CSM) of a gas of spinless
non-relativistic particles on a ring, with a scale-invariant
interaction proportional to the inverse-square chord distance between
particles. Different members of the CSM family are characterized by a
dimensionless parameter $\lambda$, where the wavefunction vanishes as
$(x_i-x_j)^{\lambda}$ as two particles approach each other. At the
coupling $\lambda$ = 1 the interaction vanishes, and the model reduces
to a gas of free spinless fermions. Early progress in obtaining
correlation functions of the model came from Dyson's work\cite{dyson}
on correlations of the eigenvalues of random matrices: at the special
couplings $\lambda$ = 1/2, 1, and 2, where the CSM is related to the
orthogonal, unitary, and symplectic random matrix ensembles. Dyson's
work led to explicit expressions\cite{sutherland} for the static
density-density correlations at $\lambda$ = 1/2 and 2, and for the
density matrix at $\lambda$ = 2. For many years, these were the only
explicitly-obtained correlations of integrable models.

The recent breakthrough derives from the work of Simons, Lee and
Al'tshuler\cite{sla} (SLA), and again has its origin in the study of
random matrix problems. Using the functional integral formalism
introduced by Efetov\cite{efetov}, SLA obtained expressions that they
initially conjectured were the {\it dynamical} density-density
correlations at $\lambda$ = 1/2 and 2, since the analogous quantity at
$\lambda$ = 1 was indeed that quantity for the free Fermi gas, and the
equal-time limit of their result coincided with the results known from
Dyson's work. Their remarkable conjecture was later verified.  A
variant of the SLA methodology was used by us\cite{hz} (HZ) to obtain
the retarded part of the bosonic single-particle Green's function at
$\lambda$ = 2, which at equal times reduces to the previously-known
expression for the density matrix. When rewritten in terms of {\it
form factors}, the SLA results become especially simple, and led to a
conjecture by one of us\cite{tata} of their generalization to all
rational coupling parameters $\lambda$. This conjecture was recently
proved using Jack polynomial techniques\cite{ha} (and was
independently made at integral values of $\lambda$, using
group-theoretical arguments\cite{min}). Jack polynomial methods were
introduced in this context by Forrester\cite{forr}, who
recently\cite{forr1} used them to prove a similar
conjecture\cite{tani} extending the HZ result to all integer
$\lambda$. Ha\cite{ha} has also extended this to arbitrary rational
$\lambda$.

While Jack polynomial methods are not restricted to particular values
of $\lambda$, they require further development. In the meantime, the
random matrix techniques provide powerful computational methods at
couplings $\lambda$ = 1/2 and 2. In this paper, we extend our earlier
result\cite{hz} by also giving the {\it advanced} part of the bosonic
single-particle Green's function at $\lambda$ = 2, and extending the
calculation to the other coupling $\lambda$ = 1/2.

The paper is organized as follows. Sect.~\ref{sec:CSM} reviews the
physics of the CSM, with particular emphasis given to the
identification of its elementary particle and hole excitations.
Anyonic particle creation and annihilation operators are defined in
Sect.~\ref{sec:psi}. Our results are stated in Sect.~\ref{sec:2}. The
two-particle Green's function defined in Sect.~\ref{sec:3} is
rewritten as a harmonic oscillator correlation function in
Sect.~\ref{sec:4}. The bulk of the paper is contained in
Sect.~\ref{sec:5}, where the supersymmetric calculation of Green's
functions is described. Many technical details have been relegated to
three appendices.

\section{The Calogero-Sutherland Model}
\label{sec:CSM}

The Calogero-Sutherland model\cite{calogero,sutherland} is a model of
a gas of non-relativistic impenetrable particles with inverse-square
interactions. It is convenient to impose periodic boundary conditions
on a ring of length $L$, although the simple results for the
correlation functions will only be given in the thermodynamic limit.
The Hamiltonian is
	\begin{equation}
	H_{\rm CS} = \sum_i {p_i^2 \over 2m } +
	\sum_{i < j } {g \over {\rm d}(x_i -x_j)^2 }
	\end{equation}
where ${\rm d}(x)$ = $(L / \pi ) \sin ( \pi x /L )$ is the chord
distance between particles on the ring. The model is fully defined by
the boundary condition on the wavefunction when two particles approach
each other:
	\[
	|\Psi | \propto |x_i - x_j|^{\lambda}, \quad | x_i - x_j |
	\rightarrow 0,
	\]
where $\lambda \ge 0 $, and
	\[
	g = {\hbar^2 \lambda (\lambda -1 ) \over m}.
	\]
Thus $\lambda \rightarrow 0 $ corresponds to the free Bose gas, and
$\lambda \rightarrow 1$ to the free (spinless) Fermi gas. (For $-1/2 <
g \le 0$, there are two different choices of $\lambda$ for each $g$.)

The energy spectra\cite{sutherland} are given by the Bethe-Ansatz
like equations, and can be viewed as adiabatic deformations of the
free-electron spectrum for $\lambda$ = 1: for $i$ = $1,\dots , N$
	\begin{equation}
	k_iL = 2\pi I_i  + \pi(\lambda -1 )\sum_{j} {\rm sign}
	(k_i - k_j) ,
	\label{BA}
	\end{equation}
where the $\{I_i\}$ are distinct integers (or for more general boundary
conditions, all $I_i - I_j$ are integral). Energy and momentum of the
eigenstate are then given by
	\begin{equation}
	E = \sum_i {(\hbar k_i)^2 \over 2m} , \quad P
	= \hbar \sum_i k_i .
	\label{E}
	\end{equation}
In the ground state, the $I_i$'s are consecutive integers, and the
$k_i$'s are equally spaced, with spacing $2\pi \lambda / L$.  If an
excited state is made by creating a ``hole'' in this ``pseudo-Fermi
sea'', by removing one of the $I_i$'s, the gap between the two
consecutive $k_i$ on either side of the hole is $2\pi (\lambda+1)/L =
(2\pi\lambda/L )(1+\lambda^{-1})$.  There is a very simple
interpretation of this: the {\it dressed charge} of the hole is
$-\lambda^{-1}$, as the extra width of the pseudo-Fermi sea associated
with the presence of the hole is $\lambda^{-1}$ of the width per
particle. (Here the bare particles are taken to have unit charge).
The other kind of excitation is to add a particle in a momentum state
outside the pseudo-Fermi sea. In this case, the charge is $+1$, the
bare charge. In the free fermion limit, $\lambda = 1$, these
identifications coincide with the usual ones for the Fermi gas.

Besides carrying fractional charge, the excitations have a natural
interpretation as particles carrying {\it fractional statistics} (see
{\it e.g.}, \cite{tani}). This is perhaps easiest to see from the
wavefunctions for the periodic model, which have the
form\cite{sutherland}
	\begin{equation}
	\Psi =  \phi (z_1,\ldots,z_N) \prod_{i<j} (z_i-z_j)^{\lambda}
	\prod_i z_i^{(J + \alpha)}
	\label{wf}
	\end{equation}
where $z_i = e^{2\pi i x_i / L}$, $\phi(\{z_i\})$ is a symmetric
polynomial that is not divisible by $z_i$ (called a Jack
polynomial\cite{forr}), $J$ is any integer, and $\alpha$ is fixed
(modulo an integer) by the choice of generalized periodic boundary
condition (the phase change of $\Psi$ when $x_i \rightarrow x_i + L
$). In the older literature\cite{sutherland}, it was usual to replace
$(z_i-z_j)^{\lambda}$ by $|z_i - z_j|^{\lambda}$ or $|z_i -
z_j|^{\lambda -1}(z_i-z_j)$, and call the bare particles bosons or
fermions, but in the light of the striking resemblance of the CSM
wavefunction to the Laughlin wavefunction\cite{laughlin}, and the
wavefunction for ``anyons'', the form (\ref{wf}) seems more
appropriate.  This strongly suggests that the bare particles of the
CSM be identified as anyons, with ``statistical parameter'' $\Theta =
\pi\lambda$.  Polychronakos\cite{polynpb} has also given arguments for
this based on scattering phase shifts.

The model for anyons is that of particles carrying both charge and
magnetic flux\cite{wilczek}. We thus can identify the bare CSM
particle as effectively carrying ``electric'' charge $Q$ and
``magnetic'' flux $\Phi$, so the statistical angle $\Theta$ is $Q\Phi
/ 2\hbar $, where
	\[
	\mbox{ particle:}\quad
	Q = e, \quad \Phi = \lambda h/e ,\quad \Theta = \pi \lambda .
	\]

What about the hole excitation?
A ``coherent'' state with a single hole has the wavefunction
	\[
	\Psi = \prod_i (z_i - Z) \Psi_0
	\]
where $\Psi_0$ is the ground state wavefunction. Here $Z$ (if
unimodular) parameterizes the position of the hole; expansion of this
wavefunction in powers of $Z$ gives its expansion in states of
definite momentum which are the one-hole eigenstates of the
Hamiltonian.  All this is essentially the same as in the Laughlin
states and the fractional quantum Hall effect (FQHE)\cite{laughlin}.
Taking over the charge-statistics results from the FQHE\cite{arovas},
leads to the identification of the charge $Q_{\rm h}$, flux $\Phi_{\rm
h}$, and statistical angle $\Theta_{\rm h}$ of the hole:
	\[
	\mbox{ hole:} \quad Q_{\rm h} = -\lambda^{-1}e , \quad
	\Phi_{\rm h} = - h/e , \quad \Theta_{\rm h} = \pi \lambda^{-1}.
	\]
Again, these identifications reduce to the trivially-expected ones when
$\lambda = 1$, but are otherwise highly non-trivial.

The CSM has Galilean invariance, and it is straightforward to compute
the effective mass of the holes from the equations (\ref{BA},\ref{E}).
(The inertial mass of the particle excitation is unchanged from the
bare value).  The hole has inertial mass $m_{\rm h}$ =
$-m\lambda^{-1}$, so that, as expected in a Galilean-invariant system,
$Q_{\rm h}/m_{\rm h} = Q/m$. This is further corroboration of the
identification of the fractional charge of the hole. Since the
particle and hole excitations have different masses, the appropriate
dynamical variable with which to describe them is their {\it
velocity}. Holes have velocities in the range $-v_s < \bar{v} < v_s$,
where $v_s$ = $\pi \lambda \hbar \rho_0 / m$ is the speed of sound
(long wavelength density fluctuations) in the model, and $ \rho_0$ =
$N/L$ is the mean density. Left-moving low-energy holes $v \approx
-v_s$ and right-moving low-energy holes $v\approx v_s$ are
continuously connected to each other. On the other hand, the particle
excitations have $|v| > v_s$, and the left- and right-moving states
are disjoint.

If an eigenstate is characterized by a set of particle excitations
with velocities $\{v_i\}$ and a set of hole excitations with
velocities $\{\bar{v}_i\}$, momentum and energy (relative to the
ground state) are given by
	\begin{eqnarray}
	P &=& \sum_i m v_i + \sum_i m_{\rm h} \bar{v}_i ,\nonumber \\
	E &=& \sum_i \half m(v_i^2-v_s^2) +
	\sum_i \half m_{\rm h} (\bar{v}_i^2-v_s^2) .
	\label{ene}
	\end{eqnarray}
Let us now consider the implication of these identifications for
the {\it form factors}: what types of excitation are produced
when the density operator
	\[
	\rho (x) = \sum_i \delta(x-x_i)
	\]
acts on the ground state? The density operator carries no charge,
and is a bosonic operator. For $\lambda = 1$, we know the answer:
the action of $\rho(x)$ on the ground state produces exactly one
hole  and one particle, with a total charge $Q + Q_{\rm h}$ = 0,
and a total flux $\Phi + \Phi_{\rm h}$ = 0.

The recent calculations of Simons, Lee, and Altshuler\cite{sla}, when
suitably interpreted, give the non-trivial answer for $\lambda= 2$ and
$\lambda = 1/2$. For $\lambda = 2$, one particle and two holes are
produced, while for $\lambda = 1/2$, two particles and one hole are
produced. This count corresponds to the number of free velocity
parameters in the form factors implied by the SLA results. As
expected, these are consistent with a total charge of zero and a total
flux of zero. The generalization of this is that {\it at rational
coupling $\lambda = p / q$, the action of the local density operator
on the CSM ground state will produce excited states with exactly $q$
particles and $p$ holes}.  A further restriction is found: in the
$\lambda $ = $1/2$ CSM, the results\cite{sla} show that both the
particles have velocities in the same direction (momenta on the same
side of momentum space with respect to the pseudo-Fermi sea). The
generalization of this is the result that all the $q$ particle
excitations have velocities in the same direction (all left-movers or
all right-movers). Thus the simplest possible excitations consistent
with charge neutrality account for the complete spectral weight of the
density-density correlation function. This is a generalization of the
free Fermi gas property of the $\lambda$ = 1 model, where just one
particle and one hole are produced by the action of the local density
operator. In a more general interacting model, there would be
contributions from two particle-hole pairs, three particle-hole pairs,
and so on. The remarkable structure of the non-trivial solutions found
at $\lambda $ = 1/2 and 2 is that the only intermediate states that
contribute are the minimally-excited ones, where the smallest number
of excitations consistent with charge neutrality are made. In this
sense, the CSM behaves like a generalized ideal gas model.

In the case of the single-particle Green's function, there are two
terms to consider: the advanced term, where a particle is added to the
system, and removed at a later time, and the retarded term where it is
removed and later restored. In these cases, the intermediate states
must again have an excitation content with the same total charge and
flux as the operator that initially acts. Thus in the advanced case,
the total charge is $e$, and the flux is $\lambda h /e$; the negatives
of these apply in the retarded case. At coupling $\lambda$ = $p / q $,
we find the expected contribution with just one particle excitation,
plus a second term with one particle in the forward direction, $q$
particles in the backwards direction, and $p$ holes. In the retarded
case, the general result found by Ha is $q-1$ particles (in the same
direction) and $p$ holes.

\section{Anyon Creation and Annihilation Operators}
\label{sec:psi}

We have reviewed the arguments in favor of the interpretation of the
CSM particles as anyons when $\lambda$ is not an integer. This is the
interpretation we adopt.

To calculate the anyonic CSM Green's functions, we shall need to
define the action of the particle creation and annihilation operators
on $N$-anyon states. Recall first how the particle annihilation
operator $\psi_x$ acts on a state of $N$ bosons ($s = +1$) or fermions
($s = -1$):
	\[
	\langle \{x_i\} | \psi_x | \Psi_N \rangle =
	\sum_{i=1}^N s^i \Psi_N(x_1,...,x_{i-1},x,x_{i+1},...,x_N).
	\]
Now, the Hilbert space of $N$-anyon CSM wavefunctions is spanned by
states of the form
	\begin{equation}
	\langle \{x_i\} | \Psi_N \rangle = S_N(\{x_i\})
	\prod_{i<j} {\rm d}(x_i-x_j)^{\lambda} ,
	\end{equation}
where $S_N(\{x_i\})$ is a symmetric function. When $\lambda$ is an
even (odd) integer, these wavefunctions are bosonic (fermionic), and
in these cases insertion into the previous equation gives (up to
normalization):
	\begin{eqnarray}
	&&\langle \{x_i\}|\psi_x|\Psi_{N+1} \rangle =
	S^{(-)}_{N,x}(\{x_i\})\prod_{i<j} {\rm d}(x_i-x_j)^{\lambda},
	\nonumber \\
	&&S^{(-)}_{N,x}(\{x_i\}) = S_{N+1} (x,\{x_i\})
	\prod_i {\rm d}(x-x_i)^{\lambda}.
	\label{annihilate}
	\end{eqnarray}
Note that the function $S^{(-)}_{N,x}$ is a symmetric function of its
$N$ arguments, not only for integral $\lambda$ but for {\it any}
$\lambda$.  Therefore, it makes sense to define the annihilation
operator $\psi_x$ by (\ref{annihilate}) at any $\lambda \ge 0$, and
this is what we do. The operator so defined properly maps the
($N$+1)-anyon Hilbert space onto the $N$-anyon Hilbert space.
Similarly, by starting from the formula
	\[
	\langle \{x_i\} | \psi_x^\dagger | \Psi_N \rangle =
	\sum_{i=1}^{N+1}
	s^i \delta(x-x_i) \Psi_N(...,x_{i-1},x_{i+1},...)
	\]
for bosons or fermions, we obtain
	\begin{eqnarray}
	&&\langle \{x_i\}|\psi_x^{\dagger}|\Psi_{N-1} \rangle =
        S^{(+)}_{N,x}(\{x_i\}) \prod_{i<j} {\rm d}(x_i-x_j)^{\lambda},
	\nonumber\\
	&&S^{(+)}_{N,x}(\{x_i\}) = \left ( \sum_{i}
	{ S_{N-1} (\{x_j, j\ne i\}) \delta (x-x_i)
	\over \prod_{j \ne i} {\rm d}(x-x_j)^{\lambda} } \right ).
	\label{create}
	\end{eqnarray}
Again, the last equations make sense, and give a valid definition
of the anyon creation operator $\psi_x^\dagger$, for any $\lambda$.

\section{Statement of result}
\label{sec:2}

Our results are as follows. Let $|0\rangle$ denote the $N$-particle
ground state. In the thermodynamic limit $N\to\infty$, $\rho_0 = N/L$
fixed, the advanced part of the single-particle Green's function ($t >
t'$),
	\[
	G_{\rm p}(x-x';t-t') = \langle 0 | \psi_x^{\vphantom{
	\dagger}}(t)\psi_{x'}^\dagger(t') | 0 \rangle ,
	\]
separates into two terms: $G_{\rm p}^{\vphantom{(0)}} = G_{\rm
p}^{(1)} + G_{\rm p}^{(2)}$. To present the result for the cases
$\lambda = 1/2, 1, 2$ simultaneously, we set $\lambda = p/q$. The
first term has the elementary form
	\[
	G_{\rm p}^{(1)}(x;t) = {m \over \pi\hbar} \int_{v_s}^\infty
	\left( v-v_s \over v+v_s \right)^{\lambda-1}\cos(kx)
	e^{-i\omega t} dv
	\]
where $k = mv/\hbar$, $\omega = m(v^2-v_s^2)/2\hbar$, and $v_s =
\pi\lambda\hbar\rho_0/m$. The second contribution arises from
fractionalization of the single particle into $q+1$ particle and
$p$ hole excitations with velocities $V := \{v_1,...,v_{q+1}\}$ and
${\bar V} := \{{\bar v}_1,...,{\bar v}_p\}$:
	\[
	G_{\rm p}^{(2)}(x;t) =
	\int \ \big| F(q+1,p,\lambda| V,{\bar V}) \big|^2 \cos(kx)
	e^{-i\omega t} dV d{\bar V} .
	\]
Here $dV = \prod_{i=1}^{q+1} dv_i$, $d{\bar V} = \prod_{j=1}^p
d{\bar v}_j$, and the integration is restricted to the domain
\end{multicols}
	\[
	-\infty < v_1 < ... < v_q < -v_s < {\bar v}_1 < ... <
	{\bar v}_p < +v_s < v_{q+1} < +\infty .
	\]
Wave number $k = P/\hbar$ and frequency $\omega = E/\hbar$ have been
given in (\ref{ene}). The ``form factor'' $|F|^2 = \big| \langle V,
{\bar V} | \psi_{x'}^\dagger|0\rangle \big|^2$ consists of two parts:
$F = F^{(0)}\times F^{(1)}$. The first one,
	\begin{equation}
	\big| F^{(0)}(n_{\rm p},n_{\rm h},\lambda|V,{\bar V})
	\big|^2 = { \prod_{i > j} (v_i-v_j)^{2\lambda} \prod_{i > j}
	({\bar v}_i-{\bar v}_j)^{2/\lambda} \over \prod_{i=1}^{n_{\rm
	p}} (v_i^2-v_s^2)^{1-\lambda} \prod_{j=1}^{n_{\rm h}}(v_s^2-{
        \bar v}_j^2)^{1-1/\lambda} \prod_{i,j} (v_i-{\bar v}_j)^2 }\ ,
	\label{jacobian}
	\end{equation}
\begin{multicols}{2}\noindent
arises in our calculation as a jacobian or measure and in this sense
is purely statistical, doing no more than measuring the phase space
available to the elementary excitations. $F^{(0)}$ is completely
determined by the root system of an underlying Lie superalgebra and is
therefore universal.  (More precisely speaking, if $n_{\rm p}$ and
$n_{\rm h}$ are the numbers of particles and holes an excitation is
composed of, the form factor of that excitation will contain a factor
$F^{(0)}(n_{\rm p},n_{\rm h},...)$.) Note also the duality symmetry
	\begin{equation}
	F^{(0)}(n_{\rm p},n_{\rm h},\lambda|V,{\bar V}) =
	F^{(0)}(n_{\rm h},n_{\rm p},\lambda^{-1}|{\bar V},V) .
	\label{duality}
	\end{equation}
The other factor is specific to the advanced part of the
single-particle Green's function and is given by
	\begin{eqnarray}
	F_2^{(1)} &=& C_2 {d \over dv}
	\left( { (v - {\bar v}_1)(v - {\bar v}_2)
	\over (v - v_1)^2 } \right)\Big|_{v=v_2} , \nonumber \\
	F_1^{(1)} &=& 0 ,			\nonumber \\
	F_{1/2}^{(1)} &=& C_{1/2} \int_{v_3}^\infty
	{dv \over \sqrt{v-v_3} } \left( {v-{\bar v}_1 \over
	\prod_{i=1}^2 \sqrt{v-v_i} } - 1 \right) .	\nonumber
	\end{eqnarray}
The normalization constants are $C_2 = \sqrt{m/8\pi\hbar}$ and
$C_{1/2} = \sqrt{m/4\pi^3\hbar}$.

Our calculation of $F_\lambda^{(1)}$ for $\lambda = 1/2, 1, 2$ is
suggestive of generalization and leads us to propose the following
conjecture for integral $\lambda = p > 1$:
	\begin{equation}
	F_\lambda^{(1)} = C_\lambda \left( {d \over dv}
	\right)^{\lambda-1} { \prod_{i=1}^\lambda (v - {\bar v}_i)
	\over (v - v_1)^\lambda } \Big|_{v = v_2} .
	\label{conjecture}
	\end{equation}
Verification of this conjecture is posed as a challenge to the
theory of Jack symmetric functions.

\section{Two-particle Green's Function}
\label{sec:3}

The objective of this paper is the calculation of the advanced part of
the single-particle Green's function. For reasons of generality and
convenience of presentation, we will set up our computational
machinery for the two-particle (or particle-hole) Green's function $(t
\ge t')$
	\[
	G_{xywz}(t-t') = \langle 0 | \psi^\dagger_y(t)
	\psi^{\vphantom{\dagger}}_x(t) \psi^\dagger_w(t')
	\psi^{\vphantom{\dagger}}_z(t') | 0 \rangle .
	\]
For $x = y$ and $w = z$, this Green's function reduces to the
dynamical density correlation function calculated in \cite{sla,ha}.
When the pair $x,w$ is taken to be remote from the pair $y,z$, we
expect $G$ to separate:
	\[
	G_{xywz}(t) \rightarrow G_{\rm p}(x-w;t) G_{\rm h}(y-z;t)
	\]
where $G_{\rm p}$ and $G_{\rm h}(x-y;t-t') = \langle 0 |
\psi^\dagger_x(t) \psi^{\vphantom{\dagger}}_y(t') | 0 \rangle$ are the
retarded and advanced parts of single-particle Green's function.

Our first task is to describe the action of the particle-hole creation
operator $\psi^\dagger_a \psi^{\vphantom{\dagger}}_b$ on the
$N$-particle ground state $| 0 \rangle$. By combining the definitions
(\ref{annihilate}, \ref{create}) we see that this action is simply
multiplication with a function of the coordinates $X :=
\{x_1,...,x_N\}$:
	\begin{eqnarray}
	&&\langle X | \psi^\dagger_a \psi^{\vphantom{\dagger}}_b |
	0\rangle = O_{ab}(X) \langle X | 0 \rangle ,	\nonumber \\
	&&O_{ab}(X) = \sum_{j=1}^N \delta(a-x_j) \prod_{k\not= j}
	{ {\rm d}^\lambda(b-x_k) \over {\rm d}^\lambda(a-x_k) } .
	\label{Oab}
	\end{eqnarray}

As it stands, the ``particle-hole function'' $O_{ab}(X)$ is not in a
form suitable for calculation by the technique used in the present
paper, which is why we make the following modification. Consider for
$\varepsilon>0$ the function $\delta_\varepsilon : {\bf R} \to {\bf
R}$ defined by ($\Re$ is the real part)
	\[
	\delta_\varepsilon(x) = \Re \ (2i\varepsilon)^\lambda (2\pi
	i)^{-1} (x+i\varepsilon)^{-\lambda} (x-i\varepsilon)^{-1} .
	\]
Since $f(0) = \lim_{\varepsilon\to 0}$ $\int_{{\bf R}}
f(x)\delta_\varepsilon(x)dx$ for any smooth function $f$,
$\delta_\varepsilon$ is a valid regularization of Dirac's delta
distribution centered at zero. We make $\delta_\varepsilon$ periodic
by replacing distance by chord distance:
	\[
	\tilde\delta_\varepsilon(x) = \Re \
	{ (2i\varepsilon)^\lambda / 2\pi i \over
	{\rm d}^\lambda(x+i\varepsilon) {\rm d}(x-i\varepsilon)} .
	\]
Inserting this into the expression for $O_{ab}(X)$, we obtain
	\begin{eqnarray}
	&&O_{ab}(X) = \Re \ { (2i\varepsilon)^\lambda / 2\pi i \over
	{\rm d}^\lambda(b-a+i\varepsilon)} \times	\nonumber \\
	&&\sum_{j=1}^N { 1 \over {\rm d}(a-x_j-i\varepsilon) }
	\prod_{k=1}^N { {\rm d}^\lambda(b-x_k+i\varepsilon) \over
	{\rm d}^\lambda(a-x_k+i\varepsilon)} ,
	\label{OabR}
	\end{eqnarray}
which differs from (\ref{Oab}) by terms that become negligible in
the limit $\varepsilon \to 0$. Introducing
	\[
	K(X,X';t) = \langle 0 | X \rangle \langle X |
	e^{-itH_{\rm CS}/\hbar} | X' \rangle \langle X' | 0 \rangle
	\]
we can now write the two-particle Green's function in the form
	\[
	G_{xywz}(t) = \int dX \int dX' \ O_{xy}(X) K(X,X';t)
	O_{wz}(X').
	\]
Here, $dX = \prod dx_i$ and the integration range is that sector
of $[0,L)^N$ where $x_i < x_{i+1}$ for $i = 1, ..., N-1$.

\section{ Mapping on a System of Harmonic Oscillators}
\label{sec:4}

It has long been known \cite{OP} that the CSM at coupling strengths
$\lambda$ = 1/2, 1, and 2 is equivalent to {\it free-particle motion}
on a symmetric space ${\rm U}(N)/{\rm O}(N)$, ${\rm U}(N)$ and ${\rm
U}(2N)/ {\rm Sp}(2N)$, respectively. More precisely speaking, for
$\lambda$ = 1/2, 1, or 2 there exists a similarity transformation
taking $H_{\rm CS}$ into the radial part of the kinetic energy
operator on the corresponding symmetric space. We will now review this
transformation for the case $\lambda = 2$.

Let ${\cal C} = 1_N \otimes i\sigma^y$ be the symplectic unit acting
on the $2N$-dimensional linear space ${\bf C}^N \otimes {\bf C}^2$.
With $g$ running through the group ${\rm U}(2N)$ we consider the set,
${\cal M}_N$, of matrices $S$ of the form
	\[
	S(g) = g {\cal C} g^{\rm T} {\cal C}^{-1} .
	\]
Such matrices are unitary and obey the self-duality constraint $S =
{\cal C} S^{\rm T} {\cal C}^{-1}$. They do not form a group. Rather,
since $S(gk) = S(g)$ for any element $k$ of the symplectic subgroup
${\rm Sp}(2N) \subset {\rm U}(2N)$ defined by the condition $k {\cal
C} k^{\rm T} = {\cal C}$, ${\cal M}_N$ is isomorphic to the coset
space ${\rm U}(2N) / {\rm Sp}(2N)$. ${\rm U}(2N)$ acts as a
transformation group on ${\cal M}_N$ by $S(g) \mapsto S(hg)$
$(h\in{\rm U}(2N))$. When endowed with its natural ${\rm
U}(2N)$-invariant metric $-\half\tr{\rm d}S{\rm d} S^\dagger$ and
probability measure $d\mu(S)$, ${\cal M}_N$ is known in random matrix
theory as Dyson's Circular Symplectic Ensemble (CSE)
\cite{dyson,Mehta}.

Dyson, in his classic paper \cite{dyson}, proved the following two
results.  (i) Any element $S \in {\cal M}_N$ has a ``polar''
decomposition $S = k e^{i\theta} k^{-1}$ where $k \in {\rm Sp}(2N)$,
and $\theta = {\rm diag} (\theta_1,...,\theta_N)\otimes 1_2$ is
diagonal. (We refer to $e^{i\theta}$ and $k$ as the ``radial'' and
``angular'' parts of $S$. Note that every eigenvalue of $S$ occurs
with multiplicity 2.) (ii) Polar decomposition takes the invariant
integral on ${\cal M}_N$ into
	\begin{eqnarray}
	&&\int_{{\cal M}_N} f(S) d\mu(S) = \nonumber \\
	&&\int_{ [0,2\pi)^N } \left( \int_{{\rm Sp}(2N)}
	f(k e^{i\theta} k^{-1})	dk \right)
	J(\theta) d\theta 		\nonumber
	\end{eqnarray}
where $d\theta = \prod_{i=1}^N d\theta_i$, and $dk$ is a suitably
normalized Haar measure for ${\rm Sp}(2N)$. The jacobian $J$ for the
transformation to polar coordinates is
	\[
	J(\theta) = \prod_{i<j} \sin^4 [(\theta_i - \theta_j ) /2].
	\]
Now, let ${\cal L}$ be the Laplacian (or minus the operator for
kinetic energy) of the Riemannian manifold ${\cal M}_N$. If ${\cal
L}_{\rm r}$ is the radial part of ${\cal L}$, we have ${\cal L} f$ =
${\cal L}_{\rm r}f$ for any radial function $f$. Using
	\[
	\tr{\rm d}S{\rm d}S^\dagger = -2 \sum_{i=1}^N
	{\rm d}\theta_i^2 + \tr [k^{-1} {\rm d}k, e^{i\theta}]^2
	\]
in conjunction with the standard coordinate expression for the
Laplacian on a Riemannian space, we easily get
	\[
	{\cal L}_{\rm r} = J(\theta)^{-1} \sum_{i=1}^N {\partial
	\over \partial\theta_i} J(\theta)
	{\partial\over\partial\theta_i} .
	\]
The differential operator ${\cal L}_{\rm r}$ is self-adjoint w.r.t.
the radial measure $J(\theta)d\theta$. It is made self-adjoint w.r.t.
the flat measure $d\theta$ by carrying out the similarity
transformation ${\cal L}_{\rm r} \to J^{1/2} {\cal L}_{\rm r}
J^{-1/2}$. With the help of the identity
	\[
	w_{ij}w_{jk} + w_{jk}w_{ki} + w_{ki}w_{ij} = 1
	\]
for $w_{ij} = \cot[(\theta_i - \theta_j)/2]$, and by making the
identification $\theta_i = 2\pi x_i/L$, we obtain
	\[
	H_{\rm CS}^{(\lambda=2)} = {(2\pi\hbar)^2 \over 2mL^2}
	\left( - J^{1/2} {\cal L}_{\rm r} J^{-1/2} +
	N(N^2-1)/3 \right).
	\]
This is the desired relation between the CSM at $\lambda = 2$
and free-particle radial motion on the symmetric space
${\cal M}_N \simeq {\rm U}(2N) /{\rm Sp}(2N)$.

For $\lambda = 2$, the function $O_{ab}(X)$ of Eq.~(\ref{OabR})
can be viewed as a radial function
	\begin{eqnarray}
	&&O_{ab}(S) = \Re \ { 2i\varepsilon^2 / \pi \over
	{\rm d}^2(b-a+i\varepsilon)} \times		\nonumber \\
	{1\over 2} &&\tr {1 \over {\rm d}(a-{L\over 2\pi i} \ln S
	-i\varepsilon)}\det{ {\rm d}(b-{L\over 2\pi i}\ln S +
	i\varepsilon) \over {\rm d}(a - {L\over 2\pi i}\ln S
	+i\varepsilon) } \nonumber
	\end{eqnarray}
on ${\cal M}_N$. Hence, the particle-hole Green's function of the CSM
can be calculated as a dynamical correlation function for
free-particle motion on ${\cal M}_N$. Note that computation of the
latter is not entirely trivial, since ${\cal M}_N$ is a curved space.

A further simplification arises in the thermodynamic limit
$N\to\infty$.  In this limit, Dyson's circular ensembles are known to
become locally equivalent to the corresponding Gaussian ensembles
\cite{dyson,Mehta}.  For the dynamical correlation functions, the
equivalence can be understood by the following argument. We go back to
our starting point and think of the eigenangles $\theta_i \in
[0,2\pi)$ of $S \in {\cal M}_N$ as a strongly correlated gas of $N$
particles with repulsive long-range forces. (These are the particles
of the $\lambda = 2$ Calogero-Sutherland model, except that now they
are confined to a ring of circumference $2\pi$ instead of $L$.)  An
important consequence of the strong correlations is that the total
system size becomes an irrelevant length scale of the gas in the
large-$N$, or high-density, limit. This comes about because the
(impenetrable) particles of the correlated gas never stray far from
their average positions and therefore do not sense the periodic
boundary conditions. There is, then, just {\it one} length scale in
this problem for large $N$, which is the mean distance, $\Delta$,
between nearest neighbors. Similarly, there is only one time scale:
the time it takes for the wave packet of a free quantum particle with
kinetic energy $-\partial^2/\partial\theta^2$ to spread over a
distance $\Delta$. The interaction, owing to its scale-invariant
$1/r^2$ short-distance form, does not introduce an extra time scale.
For these reasons, and because correlations always decay, the
dynamical correlation functions of the particle gas will go to zero on
space and time scales large compared to $\Delta$ and $\Delta^2$,
respectively. Since the spacing of $N$ particles uniformly distributed
on the unit circle is $\Delta = 2\pi/N$, we conclude that the
dynamical correlation functions are {\it fully determined by the
physics at short distances} (of order $1/N$) {\it and short times} (of
order $1/N^2$). Note that this is an indication of universality.

Returning to the equivalent problem of free-particle motion on the
symmetric space ${\cal M}_N$, we infer that the curvature of ${\cal
M}_N$ becomes an {\it irrelevant} feature in the large-$N$ limit. In
concrete terms, for the purpose of computing the dynamical
correlations near some point $p = e^{i \theta} \in {\cal M}_N$, we may
as well replace ${\cal M}_N$ by its tangent space ${\rm T}_p {\cal
M}_N$ at $p$. In this way the problem of free-particle motion on a
curved space gets simplified to free-particle motion on a flat space.
There is one complication however. Its nature, and remedy, are again
most clearly seen in the particle gas picture. If the unit circle
${\rm S}^1$ is replaced by its tangent space, which is the real line
${\bf R}$, there is nothing that holds the gas together to produce a
finite ground state density.  Fortunately, we can easily remedy the
situation by simply enclosing the gas in a box or any kind of
confining well. More precisely speaking, what we are going to do is
the following. If particles were added to the confining well with the
size of the well kept fixed, the particle density would increase.
Therefore, to have a well-defined limit when sending $N\to\infty$, we
rescale the confining potential so as to maintain a constant particle
density near the center of the well. Then, in the limit $N\to\infty$ a
translationally invariant region will form around this center, the
system will become ignorant of the confinement scale, and the
dynamical correlations will approach the universal form we are trying
to compute.

It is now clear how we should proceed. All points of a symmetric space
are equivalent, so without loss we take $p$ to be the origin, $p = e^0
=: o$.  We linearize $S = \exp iQ$ around $S = o$, with $iQ$
parameterizing the tangent space ${\rm T}_o {\cal M}_N$. From the
defining equations of ${\cal M}_N$ (i.e. $S^\dagger = S^{-1}$ and $S =
{\cal C} S^{\rm T} {\cal C}^{-1}$) we infer the linear conditions
	\begin{equation}
	Q = Q^\dagger = {\cal C} Q^{\rm T} {\cal C}^{-1} .
	\end{equation}
The correct physical dimension of length is restored by replacing $Q
\to (L/2\pi) Q$. The expression for the particle-hole function
$O_{xy}$ then takes the form
	\begin{eqnarray}
	O_{xy}(Q) &=& \Re \ { 2i\varepsilon^2 / \pi \over
	(y-x+i\varepsilon)^2} C_{xy}^\varepsilon(Q) ,	\nonumber \\
	C_{xy}^\varepsilon(Q) &=& {1\over 2} \tr{1 \over x -
	i\varepsilon - Q} \det{ y + i\varepsilon - Q
	\over x + i\varepsilon - Q } .
	\label{phfunction}
	\end{eqnarray}
To obtain the Green's functions of the CSM at $\lambda = 2$ and for $N
\to\infty$, we have to solve a one-particle problem with Hamiltonian
	\[
	H = - {\hbar^2\over 2m}{\cal L}_Q + V(Q)
	\]
where ${\cal L}_Q$ is the Laplacian of the Euclidean space ${\rm T}_o
{\cal M}_N$ with metric $\half\tr{\rm d}Q{\rm d}Q$, and $V(Q) =
V(kQk^{-1})$ is a confining radial potential with a global minimum at
$Q = 0$. If $\langle Q | 0 \rangle$ is the ground state wavefunction
of $H$ and
	\[
	K(Q,Q';t) = \langle 0 | Q \rangle \langle Q | e^{-itH/\hbar}
	| Q' \rangle \langle Q' | 0 \rangle,
	\]
we need to calculate the double integral
	\begin{equation}
	\int dQ \int dQ' \ O_{xy}(Q) K(Q,Q';t) O_{wz}(Q')
	\label{cf}
	\end{equation}
where $dQ$ is a Euclidean measure for ${\rm T}_o {\cal M}_N$. We take
the thermodynamic limit $N \to \infty$ while rescaling $V$ in such a
way that the density $\rho_0 = \langle 0|\half\tr\delta(Q)|0 \rangle$
remains fixed. By the argument given above, the expression
(\ref{cf}) with $x,y,w,z$ at a distance of order $N^0$ from the
minimum of the confining well, will then converge to the CSM
particle-hole Green's function $G_{xywz}(t)$ in this very limit.

For technical convenience we choose the potential to be harmonic:
$V(Q) = m\omega^2 \tr Q^2 / 4\hbar$. We then get a harmonic oscillator
with frequency $\omega$, ground state wavefunction $\langle Q | 0
\rangle = {\rm const}\times \exp(-m\omega\tr Q^2/4\hbar)$ and
propagator
	\begin{eqnarray}
	&&\langle Q | e^{-itH/\hbar} | Q' \rangle =
	\sqrt{ {m\omega \over 2\pi i \hbar\sin\omega t}}
	^{\ {\rm dim}{\cal M}_N}			\nonumber \\
	&&\times \exp \left[ {im\omega \over 4\hbar\sin\omega t} \tr
	\bigl((Q^2+Q'^2)\cos\omega t - 2QQ' \bigr) \right]. \nonumber
	\end{eqnarray}
When time is analytically continued to the imaginary axis, this
propagator turns into the diffusion kernel for Brownian motion in what
is called the Gaussian Symplectic Ensemble (GSE) of random matrix
theory. For future reference we note that the basic time-ordered
harmonic oscillator correlation function for self-dual sources $A$ and
$B$ is
	\begin{equation}
	\langle 0 | {\cal T} \tr BQ(t) \tr AQ(t') | 0 \rangle =
	{\hbar\over m\omega} e^{-i\omega |t-t'| } \tr AB .
	\label{oscillator1}
	\end{equation}
The average density of the GSE is known to be of semicircular
shape\cite{dyson} and in the center is given by
	\[
	\rho_0 = \sqrt{Nm\omega/\hbar\pi^2} ,
	\]
so we are going to take the thermodynamic limit holding the product
$N \omega$ fixed.

All of the above steps are easily transcribed to the cases $\lambda =
1$ and 1/2. For $\lambda = 1$ we end up with a harmonic oscillator
model of complex hermitean matrices $Q$, while for $\lambda = 1/2$ the
matrices $Q$ are real symmetric.

\section{Supersymmetric calculation of Green's functions for
$\lambda = 2$}\label{sec:5}

We have seen that the particle-hole Green's function $G_{xywz}(t)$
of the CSM at couplings $\lambda = 1/2$, 1, and 2 can be computed
as a harmonic oscillator correlation function
	\[
	G_{xywz}(t-t') = \langle 0 | {\cal T} O_{xy}(Q(t))
	O_{wz}(Q(t')) | 0 \rangle
	\]
in the thermodynamic limit. For $\lambda = 2$, the harmonic oscillator
variable $Q$ is a self-dual hermitean $2N\times 2N$ matrix, and by
using (\ref{phfunction}) we get
	\begin{eqnarray}
	&&G_{xywz}(t)={(2i\varepsilon^2/\pi)^2 \over (x-y)^2 (w-z)^2}
	\sum_{\sigma,\tau=\pm} \sigma\tau \ g^{\sigma\tau}_{xywz}(t) ,
	\nonumber \\
       	&&g^{\sigma\tau}_{xywz}(t-t') = \langle 0 | {\cal T} C^{\sigma
	\varepsilon}_{xy}(Q(t)) C^{\tau\varepsilon}_{wz}(Q(t'))
	| 0 \rangle . \label{phcorr}
	\end{eqnarray}
Similar formulas hold for $\lambda = 1/2$ and 1. Eqs.~(\ref{phcorr})
express the particle-hole Green's function as a sum of four terms.  In
Sects.~\ref{sec:5.a}-\ref{sec:5.h} we calculate $g^{+-}$, from which
$g^{-+}$ is obtained by complex conjugation followed by the time
reversal operation. The modifications necessary for $g^{++}$ and
$g^{--}$ are described in Sect.~\ref{sec:5.i}. The changes that occur
for $\lambda = 1/2$ and 1 are sketched briefly in Sect.~\ref{sec:5.j}.

\subsection{ Gaussian superintegrals}
\label{sec:5.a}

Because $C_{xy}^\varepsilon(Q)$ depends on $Q$ in a rather complicated
way, the problem of calculating the Green's function $g^{+-}$ is not
immediately tractable. As a preparatory step, we are now going to make
$Q$ appear in the argument of an exponential function.  This will
reduce the problem to calculation of Gaussian integrals.

Let $v_{k\alpha}$ and $\chi_{k\alpha}$ $(k = 1,...,N; \alpha =
\uparrow, \downarrow)$ be a set of complex commuting and anticommuting
(or Grassmann) variables.  Our starting point are the formulas
	\begin{eqnarray}
	{\rm Det} (x \pm i\varepsilon - Q)^{-1} &=& \int dv d{\bar v}
	\exp {\pm i {\bar v}(x \pm i\varepsilon - Q ) v},\nonumber \\
	{\rm Det} (y-{Q}) &=& \int d\chi d{\bar\chi}
	\exp i {\bar\chi} (y-Q) \chi,			\nonumber
	\end{eqnarray}
where $dv d{\bar v}$ and $d\chi d{\bar\chi}$ are suitably normalized
(Euclidean) measures, and
	\begin{eqnarray}
	{\bar v} (x \pm i\varepsilon - Q ) v &=& {\bar v}_{k\alpha}
	((x \pm i\varepsilon)\delta_{kl}\delta_{\alpha\beta}
	- Q_{k\alpha,l\beta}) v_{l\beta} ,		\nonumber \\
	{\bar\chi} (y-Q) \chi &=&
	{\bar\chi}_{k\alpha} (y\delta_{kl}\delta_{\alpha\beta}
	-{Q}_{k\alpha,l\beta}) \chi_{l\beta} .		\nonumber
	\end{eqnarray}
Summation over repeated indices is implied.

In later subsections a plethora of tensor products will appear, so to
avoid writing a lot of indices we are now going to switch to
basis-free notation. This we do by interpreting the integration
variables $v_{k\alpha}$ and $\chi_{k\alpha}$ as the matrix elements of
a $(1+1)\times 2N$-dimensional supermatrix $W$. More formally, we
proceed as follows. We introduce (canonical) bases
$\{e^k\}_{k=1,...,N}$, $\{e^\alpha\}_{\alpha=\uparrow,\downarrow}$,
and $\{e^B,e^F\}$, of particle-coordinate space ${\bf C}^N$, ``spin''
space ${\bf C}^2$, and super (or Boson-Fermion) space $V_{\rm bf} =
{\bf C}^{1|1}$. Elements of the dual bases are denoted by $\theta^k$
etc. We then take $W : {\bf C}^N\otimes {\bf C}^2 \to V_{\rm bf}$ to
be the linear function defined by
	\[
	W = e^B \otimes v_{k\alpha} \theta^k \otimes \theta^\alpha
	+ e^F \otimes \chi_{l\beta} \theta^l \otimes \theta^\beta .
	\]
Similarly, we define
${\tilde W} : V_{\rm bf} \to {\bf C}^N \otimes {\bf C}^2$ by
	\[
	{\tilde W} = e^k \otimes e^\alpha {\bar v}_{k\alpha} \otimes
	\theta^B + e^l \otimes e^\beta {\bar\chi}_{l\beta} \otimes
	\theta^F .
	\]
The introduction of $W$ and its adjoint ${\tilde W}$ allows us to
combine terms:
	\[
	{\bar v}_{k\alpha} Q_{k\alpha,l\beta} v_{l\beta}
	+ {\bar\chi}_{k\alpha} Q_{k\alpha,l\beta} \chi_{l\beta}
	= \tr {\tilde W} W Q^{\rm T}.
	\]
Now let $\{ E^{\mu\nu} \}_{\mu,\nu=B,F}$ be an adapted basis of
matrices in superspace that satisfy $E^{\mu\nu} e^\lambda = e^\mu
\delta^{\nu\lambda}$.  Defining $q = (x \pm i\varepsilon) E^{BB} + y
E^{FF}$ we have
	\[
	(x\pm i\varepsilon) {\bar v}_{k\alpha} v_{k\alpha} +
	y {\bar\chi}_{l\beta} \chi_{l\beta} = \tr {\tilde W} q W ,
	\]
which yields the formula
\end{multicols}
	\begin{equation}
	{\rm Det} {y-{Q} \over x\pm i\varepsilon - {Q}} = \int dW
	d{\tilde W} \exp \pm i \tr {\tilde W} (qW-WQ^{\rm T}) .
	\label{determinant}
	\end{equation}
By differentiating both sides with respect to $x$ and then setting
$y = x \pm i\varepsilon$, we obtain
	\begin{equation}
	\tr {1 \over x \pm i\varepsilon - {Q}} = \mp i \int dW
	d{\tilde W} \tr {\tilde W} E^{BB} W \ \exp \pm i \tr
	{\tilde W} ((x\pm i\varepsilon)W - W{Q}^{\rm T})  .
	\label{resolvent}
	\end{equation}

\begin{multicols}{2}
\subsection{ Implementing self-duality}
\label{sec:5.b}

Formulas (\ref{determinant}) and (\ref{resolvent}) allow us to write
the particle-hole function $C_{xy}^\varepsilon(Q(t))$ as a product of
Gaussian integrals over two supermatrices $W^{At}$ and $W^{Rt}$.
Doing the same for its complex conjugate $C_{wz}^{-\varepsilon}
(Q(t'))$, we can carry out the integration over $Q$'s in the
expression for the Green's function $g^{+-}$ by setting

	\begin{eqnarray}
	A &=& i{\tilde W^{At}} W^{At} - i{\tilde W^{Rt}} W^{Rt} ,
	\nonumber \\
	B &=& i{\tilde W^{At'}} W^{At'} - i{\tilde W^{Rt'}} W^{Rt'},
	\nonumber
	\end{eqnarray}
and using the harmonic oscillator identity
	\begin{eqnarray}
	&&\langle 0 | \exp(\tr AQ(t))\exp(\tr BQ(t')) | 0 \rangle
	\nonumber \\
	&=& \exp\half \langle 0 | \Bigl( \tr AQ(t) + \tr BQ(t')
	\Bigr)^2 | 0 \rangle .
	\label{oscillator2}
	\end{eqnarray}
Unfortunately, since $A$ and $B$ fail to be self-dual, we are not
permitted to use the simple correlator (\ref{oscillator1}) but must
work with a more complicated version thereof. As a result, the
Hubbard-Stratonovitch transformation (Sect.~\ref{sec:5.d}) is
unnecessarily complicated and the orthosymplectic symmetry of the
extremal surface emerging in the thermodynamic limit
(Sect.~\ref{sec:5.e}) is obscured. To improve the situation we make
the following modification \cite{elk}.

The idea is to enlarge the supermatrix $W$ in such a way that $P :=
{\tilde W} W$ does satisfy the self-duality condition $P = {\cal C}
P^{\rm T} {\cal C}^{-1}$. To that end, we introduce an additional
degree of freedom, called ``quasispin'', which takes two values and is
thus formally identical to a spin 1/2. Quasispin space is denoted by
$V_{\rm cd}$. (It is the cooperon-diffuson freedom of mesoscopic
conductor physics.) $W$ is extended to a supermatrix $W : {\bf C}^N
\otimes {\bf C}^2 \to V_{\rm bf}\otimes V_{\rm cd}$ by adding an
extra row index running through two values (for the two-dimensional
space $V_{\rm cd} \simeq {\bf C}^2$).

Let $W^{\rm T}$ denote the supertranspose of the supermatrix $W$.
Note that $W^{\rm TT} = \sigma W$ and ${\tilde W}^{\rm TT} = {\tilde
W} \sigma$, where $\sigma := (E^{BB}-E^{FF}) \otimes 1$ stands for
superparity. (These double transposition rules are implied by the
relation $({\tilde W}W)^{\rm T} = W^{\rm T}{\tilde W}^{\rm T}$, which
is needed for reasonable supercalculus.) The product ${\tilde W}W$
will be self-dual if we impose the conditions
	\begin{equation}
	{\tilde W} = {\cal C} W^{\rm T}\tau^{-1}, \quad
	W = \tau {\tilde W}^{\rm T} {\cal C}^{-1},
	\end{equation}
with $\tau$ some c-number matrix. Consistency requires $W^{\rm T} =
{\cal C}^{-1} {\tilde W} \tau = {{\cal C}^{-1}}^{\rm T} {\tilde
W}^{\rm TT} \tau^{\rm T}$. Since ${\cal C}^{\rm T} = - {\cal C}$, it
follows that $\tau$ must satisfy $\tau = - \sigma \tau^{\rm T}$. The
choice we make is $\tau = E^{BB} \otimes i\sigma^y + E^{FF} \otimes
1$.

The extension to quasispin space in combination with imposition of the
constraint ${\tilde W} = {\cal C} W^{\rm T} \tau^{-1}$, leaves the
number of independent components of $W$ unchanged. It is easy to see
by inspection that formulas (\ref{determinant}) and (\ref{resolvent})
remain valid if we modify $W$ in the manner described, extend $q$ to
quasispin space in the trivial way $(q \to q\otimes 1)$ and place an
extra factor of 1/2 in the exponent.

\subsection{ Integral Representation of $g^{+-}$}
\label{sec:5.c}

We now assemble the various factors needed for the Gaussian integral
representation of the Green's function $g^{+-}$ by (\ref{phfunction}),
(\ref{phcorr}), (\ref{determinant}) and (\ref{resolvent}). For the
trace of the resolvent and the ratio of determinants at time $t$, we
introduce two supermatrices $W^{At}$ and $W^{Rt}$, respectively.
Similarly, at time $t'$ we introduce a pair of supermatrices $W^{At'}$
and $W^{Rt'}$.

Since the multitude of supermatrices involved leads to lengthy
expressions, we further compactify our notation. We do this by
arranging all supermatrices vertically as a single giant supermatrix,
denoted again by $W$. In tensor-product notation,
	\begin{eqnarray}
	&&W : {\bf C}^N \otimes {\bf C}^2 \to V_{\rm bf}\otimes
	V_{\rm cd}\otimes V_{\rm ar}\otimes V_{\rm T} ,	\nonumber \\
	&&W = e^\mu \otimes e^\nu \otimes e^{\lambda} \otimes
	e^{\rho} W_{\mu\nu,i\alpha}^{\lambda\rho}
	\otimes \theta^i \otimes \theta^\alpha ,	\nonumber
	\end{eqnarray}
where $V_{\rm T} \simeq {\bf C}^2 \simeq V_{\rm ar}$ and the
subscripts ``${\rm T}$'' and ``${\rm ar}$'' refer to the two-Time and
advanced-retarded $(\pm i\varepsilon)$ structures. The new indices
take the values $\rho = t, t'$ and $\lambda = A, R$. To assemble all
factors, we define
	\begin{eqnarray}
	q = [ 	(&&x-i\varepsilon) (E^{BB}+E^{FF})
	\otimes E^{AA}\otimes E^{tt} +	\nonumber \\
		((&&x+i\varepsilon) E^{BB}
		+(y+i\varepsilon) E^{FF})\otimes
	E^{RR}\otimes E^{tt} + 		\nonumber \\
		((&&w-i\varepsilon) E^{BB}
		+(z-i\varepsilon) E^{FF})\otimes
	E^{AA}\otimes E^{t't'} + 	\nonumber \\
		(&&w+i\varepsilon) (E^{BB}+E^{FF})\otimes
	E^{RR}\otimes E^{t't'} ] \otimes 1_{\rm cd} ,	\nonumber
	\end{eqnarray}
and we introduce the matrices
	\begin{eqnarray}
	I &=& E^{BB}\otimes 1_{\rm cd}\otimes E^{AA}\otimes E^{tt},
	\ \pi_t = 1_{{\rm bf}\times{\rm cd}\times{\rm ar}}
	\otimes E^{tt} ,			\nonumber \\
	I' &=& E^{BB} \otimes 1_{\rm cd} \otimes E^{RR} \otimes
	E^{t't'} ,
	\ \pi_{t'} = 1_{{\rm bf}\times{\rm cd}\times{\rm ar}}
	\otimes E^{t't'} .				\nonumber
	\end{eqnarray}
(Here and in the following we occasionally take liberty to change
the order of the tensor products, for notational simplicity.
What is meant will always be clear from the index nomenclature
used.) With these definitions, $g^{+-}$ has the expression
	\begin{eqnarray}
	&&g^{+-}_{xywz}(t-t') = 2^{-4} \int dW d{\tilde W} \times
	\nonumber \\
	&&\tr ({\tilde W}I W) \tr ({\tilde W}I' W)
	e^{ - i \tr {\tilde W} \eta q W / 2} \times	\nonumber \\
	&&\langle 0|{\cal T}\exp {\textstyle{i\over 2}}\tr{\tilde W}
	\eta \bigl(\pi_t WQ(t) + \pi_{t'} WQ(t')\bigr) | 0 \rangle .
	\label{gintrep}
	\end{eqnarray}
A diagonal matrix $\eta$ containing elements $\pm 1$ was inserted for
convergence of the bosonic integrations, see (\ref{determinant}) and
(\ref{resolvent}). The sign of these matrix elements is fixed on the
bosonic subspace but is arbitrary on the fermionic one. The ``good''
choice for $\eta$ turns out to be \cite{efetov,vwz}
	\[
	\eta = ( E^{BB} \otimes \sigma_{\rm ar}^z + E^{FF}
	\otimes 1_{\rm ar}) \otimes 1_{{\rm cd}\times{\rm T}} .
	\]
We make this choice because it leads without further ado to the metric
of the extremal surface of Sect.~\ref{sec:5.e} being {\it Riemannian}
(as opposed to being indefinite). In contrast, if we made the more
natural choice $\eta' = \sigma^z_{\rm ar} \otimes 1_{{\rm bf} \times
{\rm cd} \times {\rm T}}$, we would eventually be forced to rotate
integration contours to achieve positive definiteness.

We add a brief discussion of symmetries. Recall that $W$ and $\tilde
W$ are related by ${\tilde W} = {\cal C} W^{\rm T} \tau^{-1}$ with
	\[
	\tau = (E^{BB} \otimes i\sigma^y_{\rm cd} + E^{FF} \otimes
	1_{\rm cd}) \otimes 1_{{\rm ar}\times {\rm T}}.
	\]
For $t = t'$ and $q$ a multiple of the unit matrix, the exponential
part of the integrand in (\ref{gintrep}) depends on $W$ only through
the combination ${\tilde W}\eta W$, which is invariant under
transformations $W \to gW$, ${\tilde W} \to {\tilde W}g^\dagger$ if
$g$ satisfies $g^\dagger\eta g = \eta$. Invariance of the constraint
${\tilde W} = {\cal C} W^{\rm T} \tau^{-1}$ requires $g^\dagger = \tau
g^{\rm T} \tau^{-1}$. The two conditions on $g$ can be combined into
the equations
	\begin{equation}
	g = (\tau\eta) {g^{-1}}^{\rm T} (\tau\eta)^{-1} =
	 \eta {g^{-1}}^\dagger \eta^{-1} .
	\label{symmetries}
	\end{equation}
The first of these defines an orthosymplectic complex Lie supergroup
${\rm Osp}(8|8)$. The second fixes a noncompact (pseudo-unitary) real
subgroup ${\rm Uosp}(4,4|8)$, which we shall denote by ${\bf G}$ for
short. ${\bf G}$ is the symmetry group of our problem and will play a
central role later.

\subsection{Hubbard-Stratonovitch transformation}
\label{sec:5.d}

Formula (\ref{gintrep}) is useful because it enables us to integrate
out the matrix variables $Q$ and $W$ and be left with a small number
(independent of $N$) of integrals to do. This is achieved as follows.

Combination of the harmonic oscillator identities (\ref{oscillator1})
and (\ref{oscillator2}) gives
	\begin{eqnarray}
	&&\langle 0 | {\cal T} \exp(\tr AQ(t))\exp(\tr BQ(t')) | 0
	\rangle \nonumber \\
	&=& \exp \ \hbar \tr (A^2 + B^2 + 2AB e^{-i\omega|t-t'|}) /
	2m\omega . \label{oscillator3}
	\end{eqnarray}
We apply this relation to (\ref{gintrep}) by identifying
	\[
	i{\tilde W}\eta\pi_t  W/2 = A, \quad
	i{\tilde W}\eta\pi_{t'} W/2 = B .
	\]
Putting $a := i\pi_t W{\tilde W}\eta/2$ and $b := i\pi_{t'} W{\tilde
W}\eta/2$ and using the cyclic invariance of the trace, we rewrite the
right-hand side of (\ref{oscillator3}) as
	\begin{equation}
	\exp \ \hbar \str (a^2 + b^2 + 2ab e^{-i\omega|t-t'|}) /
	2m\omega . \label{hubbstrat}
	\end{equation}
The symbol $\str$ denotes the supertrace. The next step is to
linearize the quadratic form $a^2 + b^2 + ...$ by the introduction of
a Hubbard-Stratonovitch field $M$. Let $M$ be a supermatrix which has
the same symmetries as the composite object $W{\tilde W}\eta$, and let
a Gaussian $M$-average be defined by
	\begin{eqnarray}
	&&\langle \bullet \rangle_{M} = \int d{M} \ \bullet \
	\exp - m\omega V(M) / 2\hbar ,	\nonumber	\\
	&&V(M) = \str ( M_{tt}^2 + M_{t't'}^2 + 2e^{i\omega|t-t'|}
	M_{tt'}^{\vphantom{2}} M_{t't}^{\vphantom{2}} ) ,
	\label{Maverage}
	\end{eqnarray}
where $M_{ij} = \pi_i M \pi_j$ for $i,j = t, t'$.
We then claim that the expression (\ref{hubbstrat}) is equal to
	\[
	\langle \exp \str M (a+b) \rangle_{M} = \langle \exp
	{\textstyle{i\over 2}}\tr{\tilde W}\eta MW \rangle_{M} .
	\]
Formal verification of this claim is the simple matter of doing a
Gaussian integral by completing the square. On a rigorous level,
however, one needs to demonstrate that the integration contours
for the bosonic-bosonic variables in $M$ can be chosen to be
uniformly convergent in $W$. This is a tricky matter which will
not be discussed here but is well understood; see Ref.~\cite{vwz},
Sect.~5.2.

Aside from being Gaussian in $W$, the last expression has the crucial
property of being invariant under unitary transformations $W \to WU$,
${\tilde W} \to U^\dagger {\tilde W}$ ($U \in {\rm Sp}(2N) \subset
{\rm U}(2N)$). This means that the particle-coordinate and spin
degrees of freedom have now been completely decoupled. Hence, doing
the Gaussian integral over the supermatrix $W$ subject to the
constraint ${\tilde W} = {\cal C} W^{\rm T} \tau^{-1}$, we get the
inverse of a superdeterminant \cite{Berezin} raised to the $N^{\rm
th}$ power:
	\[
	\int dW d{\tilde W} \ e^{ i \tr {\tilde W}
	\eta (M - q) W / 2} = \sdet^{-N} (M-q) .
	\]
By differentiating this equation with respect to the parameters
$(x-i\varepsilon)$ and $(w+i\varepsilon)$ which are contained
in $q$ and couple to $\tr {\tilde W} \eta I W$ and $\tr {\tilde W}
\eta I' W$, we can re-express $g^{+-}$ in (\ref{gintrep})
as an ${M}$-average:
	\begin{eqnarray}
	&&g^{+-}={N^2\over 4}\langle Z(M)\rangle_{M} ,\nonumber \\
	&&Z(M) = \str{I\over M-q} \str{I'\over M-q} \sdet^{-N}(M-q).
	\label{corrfunc}
	\end{eqnarray}
Note the close similarity to the starting expression (\ref{phcorr})
and (\ref{phfunction}) for $g^{+-}$: the resolvents and determinants
in $Q$-space have simply been replaced by resolvents and
superdeterminants in $M$-space.

\subsection{Thermodynamic limit}
\label{sec:5.e}

The Gaussian integral transformation from $Q$ to $M$ has reduced the
number of integration variables enormously, from order $N^2$ to order
$N^0$. The large number $N$ now appears explicitly in the integrand,
allowing us to make a saddle-point approximation which becomes exact
in the thermodynamic limit $N\to\infty$.

Recall that we take $N \to \infty$ holding the particle density
$\rho_0 = \sqrt{Nm\omega/\hbar\pi^2}$ fixed. We anticipate that the
supermatrix $M$ will be of order $N^1$ at and near the saddle point,
whereas $q$ (containing the positions $x,y,w,z$) and $|t-t'|$ are of
order $N^0$. To extract the leading large-$N$ behavior of the integral
(\ref{corrfunc}) and (\ref{Maverage}), we look for those $M$ where the
superfunction
	\begin{equation}
	\Omega(M) = \sdet^{-N}(M) \exp - m\omega \str M^2 / 2\hbar .
	\label{omega}
	\end{equation}
is maximal. By the relation
	\[
	\sdet^{-N}({M}) = \exp -N\str\ln{M} ,
	\]
such supermatrices are solutions of the saddle-point equation
$N M^{-1} + m\omega M/\hbar = 0$ or, equivalently,
	\[
	M^2 = - (N/k_0)^2
	\]
with $k_0 = \pi\rho_0$. They form supermanifolds that are invariant
under ``rotations'' ${M} \mapsto g{M}g^{-1}$ by elements $g$ of the
symmetry group ${\bf G}$.

Let us now make $M$ dimensionless by substituting for it the rescaled
expression $(iN/k_0)M$. The solution spaces of the rescaled
saddle-point equation $M^2 = 1$ can be labelled by the eigenvalues of
$M$, which are equal to either $+1$ or $-1$. It turns out that such
eigenvalues do not lie on the integration contour that is suggested by
the requirement of uniform convergence for the bosonic-bosonic
variables. However, a subset of the surfaces on which $\Omega(M)$ is
maximal can be reached by deformation of the integration contour using
Cauchy's theorem. As far as the bosonic-bosonic sector is concerned,
the signs of the eigenvalues of $M$ on such a ``reachable'' surface
are uniquely determined by the pole structure of $\Omega(M)$, which in
turn is determined by the small imaginary parts $\pm i\varepsilon$,
omitted in (\ref{omega}); see \cite{vwz}, Sect.~5.3.  As for the
eigenvalues of $M$ associated with the fermionic-fermionic sector, no
constraints from convergence and analyticity exist. In this case, the
determining agent is the limit $N\to\infty$, which singles out the
extremal surface with the largest dimension $d = d_{\rm b} - d_{\rm
f}$ (bosonic dimension minus fermionic dimension). By inspection one
finds that $d$ is largest for the surface containing the element
	\[
	\Lambda = \sigma^z_{\rm ar}
	\otimes 1_{{\rm bf}\times{\rm cd}\times{\rm T}} .
	\]
We restrict the integration in (\ref{corrfunc}) to this surface by
setting $M = g\Lambda g^{-1}$. Expanding $\sdet^{-N}(iNk_0^{-1}M-q)$
with respect to $q$ and $\exp i\omega|t-t'|$ with respect to $|t-t'|$,
and keeping only the terms that survive in the thermodynamic limit, we
find
	\begin{eqnarray}
	&&g^{+-} = -(k_0/2)^2 \int d\mu(M)
	\str(IM) \str(I'M) \nonumber \\
	&&\times\exp\left( -ik_0\str q M + i\omega_0 |t-t'| \str
	M_{tt'} M_{t't} \right)
	\label{cpm}
	\end{eqnarray}
with $\omega_0 = \hbar k_0^2 / m$.  As usual, integration over the
Gaussian fluctuations around the surface produces the uniform (or {\bf
G}-invariant) integration measure $d\mu(M)$.

\subsection{Choice of polar coordinates}
\label{sec:5.f}

What we did up to now was standard technology \cite{efetov,vwz,sla}
and obvious to the expert. Now the real work begins! Our task is to
calculate the integral (\ref{cpm}). We begin with some notational and
conceptual preparations.

The supermatrix ${M}$ has dimension $16 \times 16$, since the spaces
$V_{\rm bf}$, $V_{\rm cd}$, $V_{\rm ar}$ and $V_{\rm T}$, on the
tensor product of which ${M}$ acts, are all two-dimensional.  Recall
the definition of the symmetry group {\bf G} as the set of $16\times
16$ supermatrices $g$ that obey the conditions (\ref{symmetries}).  We
endow ${\bf G}$ with its natural metric structure given by $\str {\rm
d}g {\rm d}g^{-1}$. Let {\bf K} be the subgroup of elements $k \in
{\bf G}$ that commute with $\Lambda$.  Since ${M}= g\Lambda g^{-1}$ is
invariant under $g \mapsto gk$ $(k\in{\bf K})$, the integrand in
(\ref{cpm}) is a function on the coset space {\bf G/K}. Elements of
this coset space are denoted by $g{\bf K}$, and the uniform
integration measure on {\bf G/K} is denoted by $dg_K$.  Thus, the
integral (\ref{cpm}) is of the general form
	\begin{equation}
	\int_{\bf G/K} f(g{\bf K}) dg_K .
	\label{integral}
	\end{equation}
To calculate such an integral we need to choose a suitable system of
coordinates, or parameterization, of {\bf G/K}. In making this choice
we are guided by the general form the particle-hole Green's function
is expected to have: a good coordinate system should contain the
velocities of the elementary excitations produced by the particle-hole
operator as a subset, and it should make the integral separate into
two form factors corresponding to the final time $t$ and initial time
$t'$. In the present subsection and the one that follows, we will
describe the parameterization that meets these expectations.

Let $\tau_3$ be defined by
	\[
	\tau_3 = \pi_t - \pi_{t'} = 1_{{\rm bf} \times
	{\rm cd} \times {\rm ar}} \otimes \sigma^z_{\rm T} .
	\]
We denote by ${\bf G}_{\rm e}$ the subgroup of ${\bf G}$
whose elements commute with $\tau_3$. In other words,
${\bf G}_{\rm e}$ is that part of the symmetry group which
operates {\it at fixed time}. Furthermore, we denote by ${\bf A}$
the six-dimensional abelian subgroup of ${\bf G}$ generated by
$(i=1,2)$
	\begin{eqnarray}
	H_1 &=& E^{BB} \otimes 1_{\rm cd}
	\otimes (C_{AR}^+)_{{\rm T}\times {\rm ar}} ,	\nonumber \\
	H_2 &=& E^{BB} \otimes 1_{\rm cd}
	\otimes (C_{RA}^+)_{{\rm T}\times {\rm ar}} ,	\nonumber \\
	H_{1+2i} &=& E^{FF} \otimes E^{ii}_{\rm cd}
	\otimes (C_{AR}^-)_{{\rm T}\times {\rm ar}} ,	\nonumber \\
	H_{2+2i} &=& E^{FF} \otimes E^{ii}_{\rm cd}
	\otimes (C_{RA}^-)_{{\rm T}\times {\rm ar}}	,
	\label{abelian}
	\end{eqnarray}
with $C_{ij}^\pm = E^{tt'} \otimes E^{ij} \pm E^{t't} \otimes E^{ji}$.
These generators are {\it off-diagonal} with respect to the chosen
bases of $V_{\rm T}$ and $V_{\rm ar}$, which implies that they
anticommute with $\tau_3$ and $\Lambda$. Therefore, the elements $a$
of the group ${\bf A}$ satisfy
	\[
	a = \tau_3 a^{-1} \tau_3 = \Lambda a^{-1} \Lambda .
	\]
Technically speaking, ${\bf A}$ is a maximal abelian subgroup for the
Cartan decomposition \cite{Helgason} of ${\bf G}$ w.r.t. ${\bf K}$.

Given ${\bf G}_{\rm e}$ and ${\bf A}$, we introduce {\it nonstandard}
polar coordinates on ${\bf G/K}$ by the map
	\begin{eqnarray}
	\phi : {\bf G}_{\rm e}/{\bf M} \times {\bf A}^+ &&\to
	{\bf G/K}, \nonumber \\
	(g{\bf M},a) &&\mapsto ga{\bf K} ,	\nonumber
	\end{eqnarray}
where ${\bf M}$ is the subgroup of elements of ${\bf K}_{\rm e} = {\bf
K} \cap {\bf G}_{\rm e}$ which commute with all elements of ${\bf A}$,
and ${\bf A}^+$ is some connected open subset of ${\bf A}$ such that
$\phi$ is bijective.  The abelian factor $a$ contains the degrees of
freedom which are ``radial'' with respect to the polar coordinate
decomposition, while $g\in{\bf G}_{\rm e}$ will be parameterized by a
set of ``angular coordinates''.

By the substitution rule for superintegrals \cite{Berezin},
the polar coordinate map $\phi$ transforms the invariant integral
(\ref{integral}) into
	\begin{equation}
	\int_{{\bf A}^+} \left( \int_{{\bf G}_{\rm e}}
	f(ga{\bf K}) dg \right) J(a) da + ...
	\label{polar}
	\end{equation}
Here $dg$ is a (suitably normalized) Haar-Berezin measure for ${\bf
G}_ {\rm e}$, $da$ is a Euclidean measure on ${\bf A}$, and $J$ is the
superjacobian of the transformation. The dots indicate correction
terms (so-called ``boundary terms''), which are due to the super
nature of the integral and will be discussed later.

The jacobian $J$ is calculated in Appendix A. If we put $\ln a =
\sum_{i=1}^2 \theta_i H_i + \sum_{i=1}^4 \varphi_i H_{i+2}$ and
introduce
	\begin{eqnarray}
	v_1 &=& - v_s \cosh 2\theta_1 , \quad
	v_2 = v_s \cosh 2\theta_2 ,			\nonumber \\
	{\bar v}_1 &=& - v_s \cos 2\varphi_1 , \quad
	{\bar v}_3 = v_s \cos 2\varphi_2 , 		\nonumber \\
	{\bar v}_2 &=& - v_s \cos 2\varphi_3 , \quad
	{\bar v}_4 = v_s \cos 2\varphi_4 ,
	\label{velocities}
	\end{eqnarray}
our result for $J = |F^{(0)}|^2$ (up to normalization) is given by
(\ref{jacobian}) with $\lambda = 2$, $n_{\rm p} = 2$ and $n_{\rm h} =
4$.  We choose for ${\bf A}^+$ the fundamental positive domain defined
by
	\[
	-\infty < v_1 < -v_s < {\bar v}_1 < {\bar v}_2 < {\bar v}_3
	< {\bar v}_4 < v_s < v_2 < \infty .
	\]
We will now extract from the argument of the exponential function in
(\ref{cpm}) those parts which depend solely on the abelian factor $a$
parameterized by $\{v_i, {\bar v}_j\}$. The term multiplying $|t-t'|$
is evaluated as follows:
	\begin{eqnarray}
	&&-4\str M_{tt'} M_{t't} =
	- \str(1+\tau_3) {M} (1-\tau_3) {M}		\nonumber \\
	&=& \str \tau_3 {M} \tau_3 {M} = \str \left( \tau_3 a \Lambda
	a^{-1} \right)^2 = \str a^4\nonumber \\
	&=& \str \cosh 4 \ln a = 2\str\cosh^2 2\ln a \nonumber \\
	&=& {8\over v_s^2} \sum_i v_i^2 - {4\over v_s^2} \sum_j
	{\bar v}_j^2 =: 16 E(a)/mv_s^2 .
	\end{eqnarray}
The term $\str q{M}$ containing the coordinates $x,y,w,z$ of
the points of particle creation and annihilation is rewritten as
	\begin{eqnarray}
	&&\str qM = \str q (g a^2 \Lambda g^{-1}) 	\nonumber \\
	= &&\str q (g a^2 \Lambda g^{-1} - a^2 \Lambda)
	+ \str \Lambda q \cosh 2 \ln a .		\nonumber
	\end{eqnarray}
Defining $m_{\rm h} = - m / 2$ and introducing
	\[
	P_{\rm p} = m(v_1 + v_2) + m_{\rm h}({\bar v}_1 + {\bar v}_2),
	\quad P_{\rm h} = m_{\rm h} ({\bar v}_3 + {\bar v}_4) ,
	\]
we can write the $g$-independent term in the form
	\[
	- mv_s \str \Lambda q \cosh 2 \ln a =
	(x-w)P_{\rm p}(a) + (y-z)P_{\rm h}(a) .
	\]

\subsection{Identification of the form factor}
\label{sec:5.g}

The introduction of polar coordinates on ${\bf G/K}$ by the map $\phi$
separates the integral (\ref{cpm}) over $M$ into a (radial) integral
over the abelian group ${\bf A}$ and an (angular) integral over the
symmetry group ${\bf G}_{\rm e}$ operating at fixed time. The latter,
in turn, separates into two commuting factors, ${\bf G}_{\rm e} = {\bf
G}_{t} \times {\bf G}_ {t'}$, one for each time. Defining for $i = t,
t'$ the projections
	\[
	q_i = \pi_i q , \quad
	K_i(a) = k_0 \pi_i \Lambda \cosh 2 \ln a ,
	\]
we get
	\[ \str q (g a^2 \Lambda g^{-1}) = \sum_{i=t,t'}
	\str q_i g_i {K}_i(a) {g_i}^{-1} . \]
Hence, the integral over ${\bf G}_{\rm e}$ splits into two factors,
one for ${\bf G}_t$ and another for ${\bf G}_{t'}$. Since these are
formally identical, let us concentrate on the first one. We introduce
$\Lambda_t = \pi_t\Lambda$ and $\pi_{\rm h} = E^{FF} \otimes E^{RR}
\otimes 1_{\rm cd} \otimes E^{tt}$, $\pi_{\rm p} = \pi_t-\pi_{\rm h}$.
Then,
	\begin{eqnarray}
	q_t &=& x \pi_{\rm p} + y \pi_{\rm h} -
	i\varepsilon\Lambda_t 			\nonumber \\
	&=& \half (x+y) \pi_t + \half (x-y)(\pi_{\rm p}-\pi_{\rm h})
	-i\varepsilon\Lambda_t ,		\nonumber
	\end{eqnarray}
and, since $\str g^{-1} \pi_t g = \str \pi_t = 0$ for $g\in{\bf G}_t$,
the integral over ${\bf G}_t$ is given by
\end{multicols}
	\begin{equation}
	f_t^\varepsilon(x-y;a) = \int_{{\bf G}_t} dg \
	\str \bigl( I g K_t(a) g^{-1} \bigr)
	\exp \Big[ -i \str \bigl( {\textstyle{1\over 2}}
	(x-y)(\pi_{\rm p}-\pi_{\rm h})
	- i\varepsilon\Lambda_t \bigr) (gK_t(a)g^{-1}-K_t(a)) \Big].
	\label{formfactor}
	\end{equation}
To summarize, our complete expression for $g^{+-}$ is
	\begin{equation}
	g_{xywz}^{+-}(t) = \int_{{\bf A}^+} f_t^\varepsilon(x-y;a)
	f_0^\varepsilon(w-z;a) \exp [ -itE(a)/\hbar + i(x-w)
	P_{\rm p}(a)/\hbar + i(y-z)P_{\rm h}(a)/\hbar ] J(a) da .
	\end{equation}
\begin{multicols}{2}

Having gone through a long and indirect derivation, we are finally in
a position to interpret the mathematical structures at hand and
translate them into physics. We see that a particle-hole excitation of
the ground state of the Calogero-Sutherland model at $\lambda = 2$
fractionalizes into six elementary excitations. The velocities of
these excitations appear as the degrees of freedom of a
six-dimensional abelian group ${\bf A}$ in our treatment. There are
two particle excitations with velocities $v_1, v_2$ and four hole
excitations with velocities ${\bar v}_1,...,{\bar v}_4$.  Note that
$v_1 < -v_s$ and $v_2 > +v_s$ are on opposite sides of the
pseudo-Fermi sea. The radial functions
	\begin{eqnarray}
	P(a) &=& \half mv_s \str \pi_{t'} \Lambda\cosh 2\ln a ,
	\nonumber \\ E(a) &=& {\textstyle{1\over 8}}
	mv_s^2 \str \cosh^2 2\ln a , 		\nonumber
	\end{eqnarray}
are the total momentum and energy of the particle-hole excitation. The
equation $P = P_{\rm p} + P_{\rm h}$ decomposes the momentum into its
particle and hole components. Our expressions for $P_{\rm h}$ and
$P_{\rm p}$ show that a single hole fractionalizes into two hole
excitations (with velocities ${\bar v}_3,{\bar v}_4$) whereas a single
particle fractionalizes into two particle and two hole excitations
(with velocities $v_1,v_2;{\bar v}_1,{\bar v}_2$). Finally,
	\begin{equation}
	F(x;a) = \lim_{\varepsilon\to 0}
	{\varepsilon^2 \over \pi x^2} f_t^\varepsilon(x;a) \sqrt{J(a)}
	\label{FormFactor}
	\end{equation}
is identified as the form factor:
	\begin{eqnarray}
	\langle 0|\psi_y^\dagger \psi_x^{\vphantom{\dagger}}|a
	\rangle &=& \overline{ \langle a|\psi_x^\dagger \psi_y^{
	\vphantom{\dagger}}|0 \rangle} \nonumber \\ &=& F(x-y;a)
	e^{ixP_{\rm p}(a)/\hbar + iyP_{\rm h}(a)/\hbar} .\nonumber
	\end{eqnarray}
The last statement is not affected by the fact that $g^{+-}$ is only
one out of four contributions $g^{\sigma\tau}$ $(\sigma\tau=\pm 1)$ to
the particle-hole Green's function.  The other contributions yield the
same form factor; see Sect.~\ref{sec:5.i}.

It is possible, though by no means easy, to calculate the form factor
$F$ in closed analytic form by doing the integral in
(\ref{formfactor}).  The result is neither simple nor illuminating,
and we will not give it here. Substantial simplifications occur for
$|x-y| \gg 1/k_0$. In this limit, $F(x-y;a) \simeq F_{\rm p}(a) \times
F_{\rm h}(a)$ becomes independent of $x-y$ and separates into two
factors for a single particle and a single hole. The technical steps
of taking $|x-y| \to \infty$, $\varepsilon\to 0$, and carrying out the
final integrations, are done in Appendix B. Our result for the
single-particle form factor has been presented in Sect.~\ref{sec:2}.
The result for the single-hole form factor is the one given
previously\cite{hz}.

\subsection{Boundary terms}
\label{sec:5.h}

Berezin's proof \cite{Berezin} of the substitution rule for
superintegrals goes through without modification only for functions
with compact support, or if the integration domain has no boundary.
For polar-coordinate superintegrals such as (\ref{polar}), whose
radial domain ${\bf A}^+$ does have a boundary, one expects extra
terms (indicated by dots) to appear in general. The structure of such
``boundary terms'' was first investigated systematically by the
mathematician Rothstein \cite{Rothstein}. Unfortunately, Rothstein's
general theory is too implicit to be useful for anything but the
simplest applications. In the present problem we are dealing with the
polar-coordinate integral on a supermanifold with the rather special
structure of a symmetric space. For this case, a powerful method for
construction of all boundary terms has recently been developed
\cite{Bundschuh}. The basic idea underlying this method is sketched
for a simple example in Appendix C. Postponing the full exposition of
the method to a future publication, we are now going to state the
result obtained.

It turns out that all the boundary terms that occur for
polar-coordinate superintegrals of the type (\ref{polar}), are
associated with {\it nonintegrable singularities} of the radial
integration measure $J(a)da$. While such singularities never exist for
ordinary (i.e. nonsuper) polar-coordinate integrals, they do exist in
the present case because the superjacobian, being a superdeterminant,
puts factors both in the numerator and in the {\it denominator}. By
the nature of the polar coordinate map, the singularities all lie on
the boundary $\partial{\bf A}^+$ of the radial space ${\bf A}^+$. For
our choice of ${\bf A}^+$, nonintegrable singularities occur for (1)
$v_1 = {\bar v}_1 = {\bar v}_2 = -v_s$, and (2) ${\bar v}_3 = {\bar
v}_4 = v_2 = +v_s$. Each of these conditions defines a
three-dimensional subspace ${\bf A}_i^+$ $(i=1,2)$ of ${\bf A}^+$.
Their intersection defines a zero-dimensional subspace, which is the
unit element $a = 1$. Correspondingly, there are three boundary terms
in the present problem. To formulate them, let ${\bf M}_i$ be the
subgroup of elements of ${\bf K}_{\rm e}$ which commute with all
elements of ${\bf A}_i^+$.  Furthermore, let $J_i : {\bf A}_i^+ \to
{\bf R}$ be the reduced jacobian which is obtained by omitting from
$J$ all singular factors. With these definitions, the complete formula
for transformation to polar coordinates is
	\begin{eqnarray}
	\int_{\bf G/K} f(g{\bf K}) dg_K = &&\int_{{\bf A}^+} \left(
	\int_{{\bf G}_e/{\bf M}} f(ga{\bf K}) dg_M \right) J(a) da
	\nonumber \\ + \sum_{i=1}^2 &&\int_{{\bf A}_i^+} \left(
	\int_{{\bf G}_{\rm e}/ {\bf M}_i} f(ga{\bf K}) dg_{M_i}
	\right) J_i(a)da \nonumber \\ + &&\int_{{\bf G}_{\rm e}/{\bf
	K}_{\rm e}} f(g{\bf K}) dg_{K_{\rm e}}. \nonumber
	\end{eqnarray}

When the function $f$ is given by the integrand of (\ref{cpm}), this
formula is interpreted as follows. The first term on the right-hand
side (the ``regular'' term) corresponds to fractionalization of
$\psi_x^\dagger\psi_y^{\vphantom{\dagger}} |0\rangle$ into {\it six}
elementary excitations, as discussed above. What the appearance of two
boundary integrals with ${\rm dim}{\bf A}_i^+ = 3$ means is that there
exists the additional possibility for $\psi_x^\dagger\psi_y^{
\vphantom{\dagger}}|0\rangle$ to fractionalize into {\it three}
elementary excitations. Two of these are holes, and one is a particle.
(It is not hard to show that the regular term vanishes by symmetry for
$x = y$, so that the dynamical density correlation function is
completely determined by fractionalization into three elementary
excitations, in agreement with the findings of \cite{sla}.)  Finally,
the last boundary term, where the abelian factor $a$ is set to unity
-- i.e. no elementary excitations are present -- accounts for the fact
that the ground state is contained in $\psi_x^\dagger \psi_y^{
\vphantom{ \dagger}} | 0 \rangle$ with nonvanishing amplitude in
general. In other words, this term is equal to the product of ground
state expectation values $\langle 0 | \psi_y^\dagger \psi_x^{
\vphantom{\dagger}} | 0 \rangle\langle 0 | \psi_w^\dagger \psi_z^{
\vphantom{\dagger}} | 0 \rangle$, which determines the long-time limit
of the dynamical correlation function.

All of the boundary terms make nonvanishing contributions to $g^{+-}$,
and hence to the full particle-hole Green's function, in general.
However, when the points $x$ and $y$ (or $w$ and $z$) are removed from
each other, only one of the boundary terms survives. This is the one
that results from setting $v_1 = {\bar v}_1 = {\bar v}_2 = -v_s$. To
extract its contribution to the advanced part of the single-particle
Green's function, we proceed as in Appendix B. The reduction in
dimensionality leads to somewhat simplified calculations in comparison
with the regular term. The form factor is now a function of just a
single variable $v = v_2$. Doing the calculation, and including the
contribution from $g^{-+}$, we obtain the result for the elementary
term in the single-particle Green's function given in
Sect.~\ref{sec:2}.

\subsection{The terms $g^{++}$ and $g^{--}$}
\label{sec:5.i}

Formula (\ref{phcorr}) for the particle-hole Green's function
	\[
	G_{xywz}(t-t') = \langle 0 | \psi^\dagger_y(t)
	\psi^{\vphantom{\dagger}}_x(t) \psi^\dagger_w(t')
	\psi^{\vphantom{\dagger}}_z(t') | 0 \rangle
	\]
involves four terms: $g^{++}$, $g^{+-}$, $g^{-+}$, and $g^{--}$. We
have shown in detail how to calculate $g^{+-}$, which is related to
$g^{-+}$ by complex conjugation followed by time reversal. The
calculation of $g^{--}$ is identical except for the minor changes we
are now going to indicate.

$g^{--}$ differs from $g^{+-}$ only by $C^{-\varepsilon}_{xy}(Q(t))$
replacing $C^\varepsilon_{xy}(Q(t))$ in (\ref{phcorr}). Consequently,
to get $g^{--}$ from (\ref{corrfunc}), all we need to do is to replace
the definitions of $I$ and $q$ by
	\begin{eqnarray}
	I &=& E^{BB} \otimes 1_{\rm cd} \otimes
	E^{RR} \otimes E^{tt} ,		\nonumber \\
	q &=& [ ((x-i\varepsilon) E^{BB}
		+(y-i\varepsilon) E^{FF})\otimes
		E^{AA}\otimes E^{tt} + 	\nonumber \\
		&&(x+i\varepsilon) (E^{BB}+E^{FF}) \otimes
		E^{RR}\otimes E^{tt} + ...]\otimes 1_{\rm cd} .
	\nonumber
	\end{eqnarray}
The supertrace in the expression for the phase factor
$\exp -i k_0 \str q \Lambda \cosh 2\ln a$ changes to
	\begin{eqnarray}
	&&- v_s \str q \Lambda \cosh 2 \ln a = 2(x-w)(v_1 + v_2)
	\nonumber \\ &&-
	(y-w)({\bar v}_1+{\bar v}_2) - (x-z)({\bar v}_3+{\bar v}_4) .
	\nonumber
	\end{eqnarray}
This equation illuminates the physical content of $g^{--}$: since the
variables $v_i$ and ${\bar v}_j$ are the velocities of the elementary
excitations present in $\psi_w^\dagger(t')\psi_z(t')|0\rangle$, we see
that we are now dealing with a situation where the two holes created
by $\psi_z(t')$ are annihilated by $\psi_x(t)$ (rather than by
$\psi_y^\dagger(t)$ as before), while the two holes created by
$\psi_w^\dagger(t')$ are annihilated by $\psi_y^\dagger(t)$. It is
clear from this picture, and it is not hard to verify by direct
calculation, that $g^{--}$ vanishes in the limit $|x-y|\to\infty$,
$x-w$ and $y-z$ held fixed. The same statement applies to $g^{++}$.
Therefore, the single-particle Green's function receives no
contribution from $g^{--}$ and $g^{++}$ but is solely determined by
$g^{+-}$ and $g^{-+}$.

\subsection{The cases $\lambda =$ 1/2 and 1}
\label{sec:5.j}

A major benefit from our basis-free approach is the ease with which
the transcription to the case $\lambda = 1/2$ can be made. We now
briefly describe the essential modifications, paying no attention to
changes of normalization factors etc.

The harmonic oscillator variable $Q$ becomes a real symmetric matrix
of dimension $N \times N$, while the basic time-ordered harmonic
oscillator correlation function (\ref{oscillator1}) remains formally
the same (for symmetric $A$ and $B$). The expression for the
particle-hole function $O_{xy}$ is replaced by
	\begin{eqnarray}
	O_{xy}(Q) &=& \Re \ { \sqrt{2i\varepsilon} / 2\pi i \over
	\sqrt{y-x+i\varepsilon} } C_{xy}^\varepsilon(Q) ,\nonumber \\
	C_{xy}^\varepsilon(Q) &=& \tr{1 \over x - i\varepsilon - Q}
	\sqrt{ \det{ y + i\varepsilon - Q
	\over x + i\varepsilon - Q} } \ .		\nonumber
	\end{eqnarray}
The trace of the resolvent is treated in the same way as before.  To
implement the symmetry requirement $({\tilde W} W)^{\rm T} = {\tilde
W}W$, we take $W$ to be a linear function $W : {\bf C}^N \to V_{\rm
bf} \otimes V_{\rm cd}$ subject to the constraint
	\[
	W = \tau {\tilde W}^{\rm T}, \quad
	{\tilde W} = W^{\rm T}\tau^{-1} .
	\]
Self-consistency now implies $\tau = + \sigma\tau^{\rm T}$, and
we choose
	\[
	\tau = E^{BB}\otimes 1 + E^{FF} \otimes i\sigma^y .
	\]
Note that this is different from the case $\lambda = 2$ only by the
exchange of the bosonic-bosonic and fermionic-fermionic sectors.
Consequently, the cases $\lambda = 2$ and $\lambda = 1/2$ are
connected by a duality transformation that simply exchanges these
sectors. (See (\ref{duality}).)

To deal with the square root of the determinant, we write it as
	\[
	{ \det(y+i\varepsilon-Q) \over \sqrt{\det(y+i\varepsilon-Q)
	\det(x+i\varepsilon-Q)} } .
	\]
For the numerator we introduce a complex fermion, and for the
denominator two real bosons. (This is the same count as in the
advanced $(-i\varepsilon)$ sector.)

The Green's function $g^{+-}$ defined in (\ref{phcorr}) is still
given by (\ref{gintrep}) but with
	\begin{eqnarray}
	q = [(&&x-i\varepsilon)1_{{\rm bf}\times{\rm cd}} \otimes
	E^{AA} + \nonumber \\
	(&&y+i\varepsilon)(E^{BB}\otimes E^{22}_{\rm cd} +
	E^{FF}\otimes 1_{\rm cd})\otimes E^{RR}	\nonumber \\
	(&&x+i\varepsilon)E^{BB}\otimes E^{11}_{\rm cd} \otimes
	E^{RR} ]\otimes E^{tt} + ...	\nonumber
	\end{eqnarray}
The dots indicate analogous terms corresponding to time $t'$,
which are unchanged. Hubbard-Stratonovitch transformation and
thermodynamic limit remain the same.

The symmetry group ${\bf G}$ is defined by (\ref{symmetries}) (with
the modified expression for $\tau$), and the abelian group ${\bf A}$
appearing in the polar decomposition of ${\bf G}$ is again
six-dimensional. From the duality symmetry connecting $\lambda = 1/2$
with $\lambda = 2$, we readily see that four of the six degrees of
freedom of ${\bf A}$ are now particles and two are holes. We choose
the fundamental positive domain ${\bf A}^+$
	\[
	-\infty < v_1 < v_2 < -v_s < {\bar v}_1 < {\bar v}_2 <
	v_s < v_3 < v_4 < \infty .
	\]
The decomposition of total momentum $P$ into the components $P_{\rm
h}$ (coupling to $y-z$) and $P_{\rm p}$ (coupling to $x-w$) now reads
	\[
	P_{\rm h} = m (v_4 - 2{\bar v}_2), \quad
	P_{\rm p} = m (v_1 + v_2 + v_3 - 2{\bar v}_1).
	\]
{}From this we learn that a single particle {\it removed} from the
ground state fractionalizes into a particle (with mass $m$) and a hole
excitation (with mass $-2m$), whereas a single particle {\it added} to
the ground state fractionalizes into three particles and one hole
excitation.

The expression for the particle-hole form factor is replaced by
	\[
	F(x;a) = \lim_{\varepsilon\to 0} \sqrt{\varepsilon\over x}
	f_t^\varepsilon(x;a) \sqrt{J(a)}
	\]
where $f_t^\varepsilon$ is still given by (\ref{formfactor}) but now
	\[
	\pi_{\rm h} = (E^{BB}\otimes E^{22}_{\rm cd} +
	E^{FF} \otimes 1_{\rm cd})\otimes E^{RR}\otimes E^{tt},
	\]
by the changed formula for $q$. Evaluation of the single-particle
form factor proceeds in essentially the same way as in Appendix B.

Finally, the case $\lambda = 1$ (free fermions) is trivial in the CSM
but nontrivial in the corresponding random matrix or harmonic
oscillator formulation. Thus the balance is now reversed: rather than
solving the former using the latter, we can put this case to use by
building up our superanalytic skills from the free-fermion results.
The symmetry group of the supersymmetric integral representation of
the particle-hole Green's function at $\lambda = 1$ is a
pseudo-unitary group ${\bf G} = {\rm U}(2,2|4) \subset {\rm Gl}(4|4)$.
The dimension of a maximal abelian subgroup ${\bf A}$ for the
symmetric space ${\bf G/K}$ is four (two ``particles'' and two
``holes''). With some effort, one can show that the regular term in
the corresponding supersymmetric polar-coordinate integral vanishes.
Thus the Green's function is fully given by the boundary terms (with
two radial degrees of freedom, one particle and one hole) in this
case. This is as expected, of course, since $\psi_x^\dagger \psi_y^{
\vphantom{\dagger}}$ acting on a noninteracting Fermi sea produces
exactly one particle and one hole.

\section{Conclusion}

In conclusion, we have completed our earlier study of dynamical
correlations of the Calogero-Sutherland model by obtaining the
advanced part of the CSM single-particle Green's function at the
special couplings $\lambda = 1/2$ and 2. Our approach is based on the
relation of the CSM to radial motion in a harmonic oscillator model,
where the oscillator variable is a real symmetric $N\times N$ matrix
(self-dual hermitean $2N\times 2N$ matrix) for $\lambda = 1/2$
($\lambda = 2$). The relation becomes exact in the thermodynamic limit
of infinite particle number $N$.

To obtain the dynamical correlations of the harmonic oscillator model,
we used superanalytic methods that were developed in the context of
random matrix theory and mesoscopic conductor physics. The CSM
particle-hole Green's function for $\lambda = 1/2$ and $\lambda = 2$
was shown to be a sum of four structurally identical terms, each of
which is an integral over a Riemannian symmetric superspace ${\bf
G/K}$. The spaces ${\bf G/K}$ for both couplings are real sections of
the complex(ified) supermanifold ${\bf G}_{\bf C} / {\bf K}_{\bf C}
= {\rm Osp}(8|8) / {\rm Osp}(4|4) \times {\rm Osp}(4|4)$.  They
transform into each other by a kind of duality (or particle-hole)
transformation which exchanges the bosonic-bosonic and
fermionic-fermionic sectors. The space ${\bf G/K}$ for $\lambda = 1$
is a real section of ${\rm Gl}(4|4)/{\rm Gl}(2|2) \times {\rm
Gl}(2|2)$.

The key to further progress was the identification of a suitable polar
decomposition of ${\bf G/K}$ into a maximal totally geodesic flat
submanifold (``radial'' space) and a transverse manifold (``angular''
space). Making this decomposition corresponds to expanding the
particle-hole Green's function in a basis of energy eigenstates:
	\begin{eqnarray}
	&&\langle 0 |
	\psi^\dagger_y(t) \psi^{\vphantom{\dagger}}_x(t)
	\psi^\dagger_w(t') \psi^{\vphantom{\dagger}}_z(t')
	| 0 \rangle =	\nonumber	\\
	&&\sum_\nu \overline{\langle \nu |
	\psi^\dagger_x \psi^{\vphantom{\dagger}}_y | 0 \rangle}
	e^{-i(t-t')E_\nu / \hbar}
	\langle \nu | \psi^\dagger_w \psi^{\vphantom{\dagger}}_z
	| 0 \rangle .	\nonumber
	\end{eqnarray}
The radial coordinates of ${\bf G/K}$ were interpreted as the
velocities of the elementary particle and hole excitations which
$\psi_x^\dagger \psi_y^{\vphantom{\dagger}}|0\rangle$ decomposes into
when $N \to \infty$. Thus, doing the radial integrals amounts to
summing over $\nu$ in that limit. Moreover, the transverse manifold
was seen to be a direct product of two supermanifolds, one associated
with initial time $t'$ and the other with final time $t$. Integration
over (either) one of these yields the particle-hole form factor
$\langle \nu | \psi_x^\dagger \psi_y^{ \vphantom{\dagger}} | 0
\rangle$, see (\ref{formfactor}, \ref{FormFactor}).

The radial space of ${\bf G/K}$ has both compact and noncompact
degrees of freedom. On making the identification with elementary
excitations of the CSM, the former translate into holes, and the
latter into particles. For $\lambda = 1/2$ ($\lambda = 2$) we find two
(four) holes and four (two) particles. The generalization of this to
rational $\lambda = p/q$ is that there will be $2p$ holes and $2q$
particles in $\psi_x^\dagger\psi_y^{ \vphantom{\dagger}}|0\rangle$ in
general. However, this not the whole story yet. When the substitution
rule is applied to a superintegral $\int_{\bf G/K} f(\xi) d\xi$, it is
found that such an integral expressed in polar coordinates, must be
corrected by the addition of several lower-dimensional integrals (the
``boundary'' terms). These corrections are a novel feature of
superintegration, and are needed to restore invariance of the integral
under translations $\xi \mapsto g \cdot \xi$ $(g\in{\bf G})$, which is
lost when the naive substitution rule involving only the ``regular''
term is assumed. In our case there are three distinct boundary terms.
For $\lambda = 1/2$ ($\lambda = 2$), two of them have a radial
integration domain which is three-dimensional, with two (one) of the
radial degrees of freedom being noncompact, i.e. particle-like, and
one (two) compact or hole-like. The remaining boundary term accounts
for the ground state component in $\psi_x^\dagger\psi_y^{\vphantom{
\dagger}}| 0\rangle$. For $x = y$ the regular term in the
polar-coordinate integral vanishes, and we recover the result that the
dynamical density correlations at $\lambda = p/q$ are exhausted by
contributions from $p$ holes and $q$ particles. The single-particle
form factors were extracted from $\langle \nu | \psi_x^\dagger
\psi_y^{ \vphantom{\dagger}}| 0\rangle$ by separation of points
$(|x-y|\to\infty)$.

The methods described in this paper readily extend to many-particle
Green's functions. By following the steps of
Sects.~\ref{sec:3}--\ref{sec:5} we can, in principle, derive an exact
expression for the form factor of any excited state $|\nu\rangle$
which is produced by several actions of $\psi$ and $\psi^\dagger$ on
the ground state. Such a form factor always separates into two parts.
The first one, $F^{(0)}$, is universal and is given by
(\ref{jacobian}). The second one, $F^{(1)}$, is given in the form of
an integral representation analogous to
(\ref{formfactor},\ref{FormFactor}).  For example, for $\lambda = 2$
we get

	\begin{eqnarray}
	&&\langle \nu | \psi_{x_1} ... \psi_{x_n} | 0 \rangle
	= F^{(0)}(\nu) F^{(1)}(x_1 ... x_n | \nu) ,\nonumber	\\
	&&F^{(1)}(x_1 ... x_n | \nu) = \prod_{i < j}(x_i-x_j)^2\int_{
	{\rm O}(2n)} e^{{m_{\rm h}\over i\hbar}\tr \nu gXg^{-1}} dg ,
	\nonumber	\\
	&&\nu = {\rm diag}(\bar v_1{\vphantom{'}},\bar v_1',...,
	\bar v_n{\vphantom{'}},\bar v_n'), \
	X = {\rm diag}(x_1,x_1,...,x_n,x_n)
	\nonumber
	\end{eqnarray}
where $dg$ is a Haar measure for ${\rm O}(2n)$. (Actually, because of
the twofold degeneracies of $X$, it is sufficient to integrate over
${\rm O}(2n)/{\rm O}(2)^n$.)

As a final remark, we are intrigued by the finding that the $p$ holes
and $q$ particles excited by the CSM density operator, can be
interpreted as radial degrees of freedom of a symmetric supermanifold
${\bf G/K}$ for $\lambda = 1/2, 1$, and 2. We wonder whether there
could exist a ``quantum'' deformation (with deformation parameter
$\lambda = p/q$) of the symmetric supermanifold, such that the maximal
``torus'' of the deformed manifold had $p$ compact and $q$ noncompact
directions. If such a deformation existed, the treatment of our paper
might carry over to more general values of $\lambda$, and the problem
of calculating correlation functions for the Laughlin wavefunction in
one dimension and in the thermodynamic limit, might be solved.

MRZ acknowledges partial support by the Deutsche
Forschungsgemeinschaft, SFB 341 K\"oln-Aachen-J\"ulich. FDMH was
supported in part by the National Science Foundation grant
DMT92-24077.

\section*{Appendix A: Calculation of a Jacobian}

We are going to calculate the jacobian $J$ associated with the map
$\phi : {\bf G}_{\rm e}/{\bf M} \times {\bf A}^+ \to {\bf G/K}$,
$(g{\bf M},a) \mapsto ga{\bf K}$. According to its definition by the
substitution rule for superintegrals \cite{Berezin}, $J$ is the
superdeterminant of the supermatrix that expresses the differential of
$\phi$ in an orthonormal basis of the tangent spaces of ${\bf G}_{\rm
e}/{\bf M} \times {\bf A}^+$ and ${\bf G/K}$. Let ${\cal G}$, ${\cal
K}$, ${\cal A}$, ${\cal G}_{\rm e}$, and ${\cal M}$ be the Lie
algebras of ${\bf G}$, ${\bf K}$, ${\bf A}$, ${\bf G}_{\rm e}$ and
${\bf M}$, respectively, and let ${\cal P}$ be the tangent space of
${\bf G/K}$ at the origin $1 \cdot {\bf K}$. Since the invariant
measure on ${\bf G/K}$ is not changed by left translation, $J(a)$
equals the superdeterminant of the linear transformation $T_a : ({\cal
G}_{\rm e}-{\cal M}) \times {\cal A} \to {\cal P}$ given by
	\begin{eqnarray}
	T_a(Z,H) &=& (ga)^{-1} {d \over ds} g e^{sZ} a e^{sH}
	{\bf K} \Big|_{s = 0}				\nonumber \\
	&=& H + \left( {\rm Ad}(a^{-1}) Z \right)_{\cal P} ,\nonumber
	\end{eqnarray}
where ${\rm Ad}(a)Z = aZa^{-1}$ denotes the adjoint action, and
$Z_{\cal P}$ means the ${\cal P}$-component of $Z \in {\cal G}$
with respect to the orthogonal decomposition ${\cal G} = {\cal K}
+ {\cal P}$.

To calculate $J(a) = \sdet T_a$, we further decompose ${\cal K}$ and
${\cal P}$ into their even parts, ${\cal K}_{\rm e}$ and ${\cal
P}_{\rm e}$, and odd parts, ${\cal K}_{\rm o}$ and ${\cal P}_{\rm o}$,
with respect to the involutory automorphism $Z \mapsto \tau_3 Z
\tau_3$.  Since the elements of ${\cal A} \subset {\cal P}_{\rm o}$
anticommute with both $\Lambda$ and $\tau_3$, the following
commutation relations are obvious:
	\begin{eqnarray}
	&&[ {\cal A} , {\cal P}_{\rm e} ] \subset {\cal K}_{\rm o},
	\quad [ {\cal A} , {\cal K}_{\rm e} - {\cal M} ] \subset
	{\cal P}_{\rm o} - {\cal A} , \nonumber \\
	&&[ {\cal A} , {\cal K}_{\rm o} ] \subset {\cal P}_{\rm e} ,
	\quad [ {\cal A} , {\cal P}_{\rm o} - {\cal A} ] \subset
	{\cal K}_{\rm e} - {\cal M} . \nonumber
	\end{eqnarray}
{}From these we see that, if $Z = X + Y$ is the decomposition of $Z$
by ${\cal G}_{\rm e} - {\cal M} = {\cal P}_{\rm e} + ({\cal K}_{\rm e}
- {\cal M})$, then
	\[
	T_a(X+Y,H) = H 	+ \cosh \ad (\ln a) \ X
			- \sinh \ad (\ln a) \ Y
	\]
where we have introduced $\ad(\ln a) X := [\ln a,X]$ (commutator)
and used the relation $\Ad(a) = \exp \ad(\ln a)$. It follows that
	\begin{eqnarray}
	J(a) = &&\sdet \left( \cosh \ad (\ln a) \Big|_{{\cal P}_{\rm
	e} \to {\cal P}_{\rm e}} \right)\times\nonumber \\ &&\sdet
	\left( \sinh \ad (\ln a) \Big|_{{\cal K}_{\rm e}-{\cal M}
	\to {\cal P}_{\rm o}-{\cal A}} \right) .	\nonumber
	\end{eqnarray}

The superdeterminant of a supermatrix equals the product of its
bosonic eigenvalues divided by the product of its fermionic ones. We
thus need the eigenvalues of the linear operators
	\begin{eqnarray}
	&&\ad(\ln a) : \ {\cal K}_{\rm o} + {\cal P}_{\rm e} \to
	{\cal P}_{\rm e} + {\cal K}_{\rm o} , 		\nonumber \\
	&&\ad(\ln a) : \ {\cal K}_{\rm e} + {\cal P}_{\rm o} \to
	{\cal P}_{\rm o} + {\cal K}_{\rm e} . \nonumber
	\end{eqnarray}
Because $\ad(\ln a)$ is block off-diagonal, these occur in pairs with
opposite sign. Let us denote a set of positive eigenvalues by
$\Delta_{\rm o}^+$ and $\Delta_{\rm e}^+$, respectively. Furthermore,
let $|m_\alpha|$ denote the multiplicity of eigenvalue
$\alpha$, and put ${\rm sign}(m_\alpha) = +1$ ($-1$) for $\alpha$
bosonic (fermionic). Then the expression for $J$ becomes
	\[
	J(a) =
	\prod_{\alpha\in\Delta_{\rm e}^+}
	\Bigl( \sinh \alpha(\ln a) \Bigr)^{m_\alpha}
	\prod_{\beta\in\Delta_{\rm o}^+}
	\Bigl( \cosh \beta(\ln a) \Bigr)^{m_\beta} .
	\]
Note that the derivation of this formula has used no more than the
generic geometric structures underlying any symmetric space ${\bf
G/K}$ with an involution $\tau_3$. It is therefore valid for all
of the cases $\lambda = 1/2$, $\lambda = 1$ and $\lambda = 2$.

Calculation of eigenvalues $\alpha$ (also called roots) and
multiplicities $m_{\alpha}$ is a standard exercise in linear algebra.
Adopting the parametrization $\ln a = \sum_{i=1}^2 \theta_i H_i +
\sum_{i=1}^4 \varphi_i H_{i+2}$ with $H_i$ given in (\ref{abelian}) we
find the following eigenvalues (resp. multiplicities) for $\lambda =
2$:
	\begin{eqnarray}
	\Delta_{\rm o}^+ : \ &&\theta_1 \pm \theta_2 \ (+4) ,
	\nonumber \\ && i\varphi_k \pm i\varphi_l \ (+1) \quad
	(k < l ; \ k + l \ {\rm odd}) ,	\nonumber \\ && \theta_k \pm
	i\varphi_l \ (-2) \quad (k+l\ {\rm odd}),\nonumber \\
	\Delta_{\rm e}^+ : \ &&2\theta_k \ (+3) \quad (k = 1, 2),
	\nonumber \\ && i\varphi_k \pm i\varphi_l \ (+1) \quad (k < l
	; \ k + l \ {\rm even}) , \nonumber \\ && \theta_k \pm
	i\varphi_l \ (-2) \quad (k + l\ {\rm even}).\nonumber
	\end{eqnarray}
The roots $\alpha$ for $\lambda = 1/2$ are obtained from this
by a simple duality transformation $\alpha \to i\alpha$.

The jacobian $J$ is easily converted to the form given in
(\ref{jacobian}), $J \propto |F^{(0)}|^2$, by using the trigonometric
identities
	\begin{eqnarray}
	2 \sinh(A+B)\sinh(A-B) &=& \cosh(2A)-\cosh(2B) ,\nonumber \\
	2 \cosh(A+B)\cosh(A-B) &=& \cosh(2A)+\cosh(2B) ,\nonumber
	\end{eqnarray}
and recalling the substitutions (\ref{velocities}). The duality
symmetry (\ref{duality}) relating $\lambda = 2$ with $\lambda = 1/2$
is a consequence of the duality of the corresponding root systems
$(\alpha \leftrightarrow i\alpha)$.

\section*{Appendix B: Calculation of the single-particle form factor}

We are going to extract the single-particle form factor $F_{\rm p}(a)
= \langle 0 | \psi_x | a \rangle \exp -ixP_{\rm p}(a)/\hbar$ from the
particle-hole form factor $F(x-y;a)$ by taking $|x-y|\to\infty$ in
(\ref{formfactor}) and (\ref{FormFactor}). In this limit, the
integrals over some of the degrees of freedom of $G_t \simeq {\rm
Uosp}(2,2|4)$ can be done by stationary phase. We begin by isolating
these degrees of freedom.

First of all, we make the transformation $g \mapsto g^{-1}$ which
turns the term coupling to $x-y$ into $\str K_t(a) g(\pi_{\rm p}-
\pi_{\rm h})g^{-1}$ and, being an isometry, leaves $dg$ unchanged. We
abbreviate $\pi_{\rm p} - \pi_{\rm h} =: D$. Next, we decompose ${\bf
G}_t$ by ${\bf G}_t \simeq ({\bf G}_t / {\bf H}) \times {\bf H}$ where
	\[
	{\bf H} = \{ h \in {\bf G}_t | h D h^{-1} = D \} ,
	\]
and we write $g \in {\bf G}_t$ as $g = s h$, accordingly. The factor
$h \in {\bf H}$ drops out of the expression $gDg^{-1} = sDs^{-1}$.
Moreover, the measures match under factorization, i.e. $dg$ turns into
$ds_H dh$, with $ds_H$ ($dh$) being the invariant measure on ${\bf
G}_t / {\bf H}$ (${\bf H}$). Now, because the phase of
	\[
	A(s) = \exp - {\textstyle {i\over 2}} (x-y) \str K_t(a)
	(sDs^{-1}-D)
	\]
oscillates rapidly for $|x-y| \gg 1/k_0$, the integral over ${\bf
G}_t/ {\bf H}$ can be done by using the stationary phase
approximation. Setting $s = \exp X$ and doing the Gaussian integral
that results upon expansion to lowest nonvanishing order in $X$, we
get
	\begin{eqnarray}
	&&(x-y)^{-2} \int_{{\bf G}_t/{\bf H}} A(s) ds_H
	\buildrel {|x-y|\to\infty} \over \longrightarrow \nonumber \\
	&&\prod_{i=3}^4 { (v_2-{\bar v}_{i})({\bar v}_{i}-v_1) \over
	({\bar v}_{i}-{\bar v}_1)^{1/2} ({\bar v}_{i}-{\bar v}_2)^{1/2}
	} \nonumber
	\end{eqnarray}
This expression precisely cancels that part of the phase space factor
$\sqrt{J}$ which connects the velocities of the ``particle set''
$(v_1,v_2,{\bar v}_1,{\bar v}_2)$ with those of the ``hole set''
$({\bar v}_3,{\bar v}_4)$, so that $\sqrt{J}$ gets replaced by the
product $\sqrt{J_{\rm p}} \times \sqrt{J_{\rm h}} = F_{\rm p}^{(0)}
\times F_{\rm h}^{(0)}$ (Eq.~(\ref{jacobian})). By the particle-hole
structure of $D = \pi_{\rm p} - \pi_{\rm h}$, the group ${\bf H}$
decomposes as
	\[
	{\bf H} \simeq {\bf H}_{\rm p} \times {\bf H}_{\rm h} =
	{\rm Uosp}(2,2|2) \times {\rm O}(2) .
	\]
Hence, the form factor separates as expected:
$F \to F_{\rm p} \times F_{\rm h}$, and we have
$F_{\rm p} = F_{\rm p}^{(0)} \times F_{\rm p}^{(1)}$,
$F_{\rm h} = F_{\rm h}^{(0)} \times F_{\rm h}^{(1)}$.
The factor pertaining to the single hole is trivial,
	\[
	F_{\rm h}^{(1)}(a) = \int_{{\bf H}_{\rm h}} dh={\rm const},
	\]
while the single-particle form factor $F_{\rm p}^{(1)}(a)$ equals
	\[
	\varepsilon^2 \int_{{\bf H}_{\rm p}} dh
	\str(K_t(a)hIh^{-1}) \ e^{-\varepsilon \str K_t(a)
	(h\Lambda_t h^{-1}-\Lambda_t)} ,
	\]
where a normalization constant has been omitted.

The remaining task is to carry out the integration over the group
${\bf H}_{\rm p}$. Simplifying the notation by writing $K$ for
$K_t(a)$ and omitting a term $\varepsilon\str\Lambda_t K$, which
vanishes in the limit $\varepsilon\to 0$, we encounter the integral
	\[
	\varepsilon^2 \int_{{\bf H}_{\rm p}} dh
	\str(K hIh^{-1})
	\exp -\varepsilon\str K h \Lambda_t h^{-1} . \nonumber
	\]
We will do this integral in several steps. The first one is to factor
${\bf H}_{\rm p}$ by ${\bf H}_{\rm p} \simeq ({\bf H}_{\rm p}/{\bf H}_
{\Lambda}) \times {\bf H}_\Lambda$ with ${\bf H}_\Lambda \simeq {\rm
Uosp}(2|2) \times {\rm Sp}(2)$ being the subgroup of ${\bf H}_{\rm p}$
that fixes $\Lambda_t$ w.r.t. the action $\Lambda_t \mapsto h
\Lambda_t h^{-1}$. Again, the invariant measures match. ${\bf
H}_\Lambda$ is a maximal compact subgroup of ${\bf H}_{\rm p}$, so
${\bf H}_{\rm p} / {\bf H}_\Lambda$ is a (super-)symmetric space of
the noncompact type. The integrand is constant on ${\rm Sp}(2)$ and
depends on ${\rm Uosp}(2|2)$ only through the combination $hIh^{-1}$
in the pre-exponential factor. Therefore, the integral over ${\bf
H}_\Lambda$ is trivial and can be done by ``orthogonality''. (By
orthogonality we mean the relation
	\[
	\int_G dU \ {\cal D}(U)_{\alpha\beta}
	{\cal D}(U^{-1})_{\gamma\delta} =
	c_0 (-1)^{|\beta|} \delta_{\alpha\delta}
	\delta_{\beta\gamma} ,
	\]
valid for any irreducible representation ${\cal D}$ of a compact
supergroup $G$.) Thus, with $\pi_+$ being the ${\rm Uosp}(2|2)$ scalar
$\pi_+ = 1_{{\rm bf}\times{\rm cd}}\otimes E^{AA}\otimes E^{tt}$,
integration over ${\bf H}_\Lambda$ simply causes the substitution
$hIh^{-1} \mapsto c_0 h \pi_+ h^{-1}$, followed by restriction of the
integration domain from ${\bf H}_{\rm p}$ to ${\bf H}_{\rm p}/{\bf
H}_\Lambda$.

To do the integral over the latter, we exploit the simplifications
that arise from the limit $\varepsilon\to 0$. What must happen is that
the integral over the noncompact manifold ${\bf H}_{\rm p}/{\bf
H}_\Lambda$ becomes singular as $\varepsilon^{-2}$ for $\varepsilon\to
0$, thereby cancelling the prefactor $\varepsilon^2$. This singular
behavior implies that the integral is dominated by the contributions
from the asymptotic region on ${\bf H}_{\rm p}/{\bf H}_\Lambda$.  We
can therefore replace the integral by its {\it contraction} to the
asymptotic region. Using the relation $h (\Lambda_t - 2\pi_+) h^{-1} =
\Lambda_t - 2 \pi_+$ we see that $\pi_+$ may be replaced by
$\Lambda_t/2$ in the integrand with an error that becomes negligible
in the limit $\varepsilon\to 0$. We set $Q := \varepsilon h\Lambda_t
h^{-1}$, which satisfies $Q^2 = \varepsilon^2 \pi_t$, so that $Q^2 =
0$ on the contracted (or ``light-cone'') manifold that emerges in the
limit $\varepsilon\to 0$. The invariant measure $\varepsilon \cdot
dh_{{\bf H}_{\rm p}/{\bf H}_\Lambda}$ can be shown to contract to the
invariant measure $dQ$ on the light-cone $Q^2 = 0$. Hence, the
integral giving the single-particle form factor $F_{\rm p}^{(1)}$
reduces to
	\begin{eqnarray}
	&&\int_{Q^2 = 0} dQ \ \str KQ \ e^{-\str KQ} \nonumber \\
	= &&-{d\over dt}\Big|_{t=1}\int_{Q^2=0}dQ \ e^{-t \str KQ}.
	\label{Bformf}
	\end{eqnarray}
To do the integral over $Q$, we write
	\[
	Q = \sum_{i,j=A,R} (Q_{ij})_{{\rm bf}\times{\rm cd}}
	\otimes E_{\rm ar}^{ij} \otimes E^{tt}
	\]
and solve the constraint $Q^2 = 0$ by putting
	\begin{eqnarray}
	Q_{AA} &=& ZZ^\dagger ,\quad Q_{AR}= -Z \sqrt{Z^\dagger Z},
	\nonumber \\ Q_{RA} &=& \sqrt{Z^\dagger Z} \ Z^\dagger ,
	\quad Q_{RR} = - Z^\dagger Z , 	\nonumber
	\end{eqnarray}
with
	\[
	Z = E^{BB} \otimes
	\pmatrix{	z_1 	&\bar z_2 \cr
			-z_2	&\bar z_1 \cr}_{\rm cd} +
	E^{FB} \otimes
	\pmatrix{ 	\zeta_1 &{\bar\zeta}_1 \cr
			\zeta_2	&{\bar\zeta}_2 \cr}_{\rm cd} .
	\]
The invariant measure $dQ$ in these variables is the Euclidean
measure $dZ dZ^\dagger = dz_1 d\bar z_1 dz_2 d\bar z_2
\partial_{\zeta_1}\partial_{\bar\zeta_1}\partial_{\zeta_2}
\partial_{\bar\zeta_2}$ multiplied by $\str Z Z^\dagger$.
Upon making the decomposition
	\[
	K = \sum_{i=A,R} (K_i)_{{\rm bf}\times{\rm cd}}
	\otimes E_{\rm ar}^{ii} \otimes E^{tt} ,
	\]
we get $\str KQ = \str Z^\dagger (K_A Z - Z K_R)$, and our
integral becomes
	\[
	- {d\over dt} \Big|_{t=1} \int dZ dZ^\dagger
	\str Z Z^\dagger \ e^{-t \str Z^\dagger (K_A Z - Z K_R)} .
	\]
By the property of perfect grading (i.e. equal number of bosonic and
fermionic degrees of freedom), the Euclidean measure $dZ dZ^\dagger$
is invariant under scale transformations $Z \to Z / \sqrt{t}$.
Therefore, carrying out this transformation and then taking the
derivative at $t = 1$, we obtain
	\[
	\int dZ dZ^\dagger \ \str Z Z^\dagger \
	e^{- \str Z^\dagger (K_A Z - Z K_R)} .
	\]
This is a Gaussian integral which is easy to calculate. The
result is the one given in Sect.~\ref{sec:2}.

We conclude this appendix by sketching the evidence in favor of the
conjecture (\ref{conjecture}). When proper meaning is given to the
quantities $K$ and $Q$, the single-particle form factor $F_{\rm
p}^{(1)}$ can be written in the form (\ref{Bformf}) for all the cases
$(\lambda = 1/2, 1, 2)$ accessible to our method. The invariant
measure $dQ$ takes the general form $dZ dZ^\dagger (\str Z
Z^\dagger)^{\lambda-1}$. By doing the same manipulations as before, we
obtain
	\[
	F_{\rm p}^{(1)} = (\lambda-1) \int dZ dZ^\dagger
	(\str Z Z^\dagger)^{\lambda-1} e^{\str Z^\dagger (Z K_R - K_A Z)} .
	\]
For integral $\lambda$, the pre-exponential factor $(\str Z
Z^\dagger)^{\lambda-1}$ can be generated by shifting $K_R \to K_R +
(k_0 v/v_s) \cdot 1$ and taking $\lambda-1$ derivatives w.r.t. $v$ at
$v = 0$. If we switch to the notation of Sect.~\ref{sec:2}, the
resulting Gaussian integral equals
	\begin{eqnarray}
	\int dZ dZ^\dagger \
	&&e^{-\str Z^\dagger(K_A Z - Z K_R - Z k_0 v/v_s)}\nonumber \\
	= &&{\prod_{i=1}^p (v_{q+1}+v-{\bar v}_i) \over \prod_{j=1}^q
	(v_{q+1}+v-v_j)^\lambda}			\nonumber
	\end{eqnarray}
in all cases. This is why we believe that (\ref{conjecture}) holds
true in general.

\section*{Appendix C: Construction of Boundary Terms}

We sketch our general procedure for constructing the boundary terms
that are associated with polar-coordinate superintegrals on
supersymmetric spaces.  Consider the simple example of the Euclidean
superplane ${\rm E}_{2|2}$ with complex supercoordinates $(z,\zeta)$
and invariant Berezin integral
	\[
	\int_{{\rm E}_{2|2}} f d\mu =
	(4\pi i)^{-1} \int_{\bf C} \Big( \partial_\zeta
	\partial_{\bar\zeta} f(z,\bar z,\zeta,\bar\zeta) \Bigr)
	dz d\bar z .
	\]
The metric of ${\rm E}_{2|2}$, ${\rm d}\bar z{\rm d}z + {\rm
d}\bar\zeta {\rm d}\zeta$, is preserved by transformations
	\[
	\pmatrix{z\cr \zeta\cr} \mapsto k
	\pmatrix{z\cr \zeta\cr}
	\]
if $k \in {\rm U}(1|1)$.  We introduce polar coordinates $(k,r) =
\phi^{-1}(z,\bar z,\zeta,\bar\zeta)$ via the diffeomorphism $\phi :
{\rm U}(1|1)/{\rm U}(1) \times {\bf R}^+ \to {\rm E}_{2|2}$ defined
by
	\[
	\pmatrix{z\cr \zeta\cr} = k \pmatrix{r\cr 0\cr} .
	\]
The substitution rule for superintegrals yields
	\[
	\int_{{\rm E}_{2|2}} f d\mu =
	\int_{{\bf R}^+} \left( \int_{{\rm U}(1|1)} (f \circ \phi)
	(k,r) dk \right) r^{-1} dr + {\cal R}[f]
	\]
where we have extended the angular integration to the group ${\rm
U}(1|1)$ for convenience. The normalization constant $1/4\pi i$ has
been absorbed into the Haar-Berezin measure $dk$. ${\cal R}[f]$ is a
correction term which is due to the radial space ${\bf R}^+$ having a
boundary. To construct it, we exploit the translational invariance of
the Berezin integral on ${\rm E}_{2|2}$ as follows. We write $X :=
(z,\bar z,\zeta, \bar\zeta)$ for short and denote the scalar product
of two vectors $X,X'$ by
	\[
	\langle X , X' \rangle = \Re \ (\bar z z' + \bar\zeta \zeta').
	\]
If $X_0 = (z_0,\bar z_0,\zeta_0,\bar\zeta_0)$ is some set of
parameters, we define a directional derivative $D_0 f$ by
	\[
	D_0 f = \langle X_0 , {\rm grad}f \rangle .
	\]
{}From $\int_{{\rm E}_{2|2}} (D_0 f) d\mu = 0$ we deduce
	\[
	{\cal R}[D_0 f] = - \int_{{\bf R}^+} \left( \int_{{\rm U}
	(1|1)} (D_0 f \circ \phi)(k,r) dk \right) r^{-1} dr .
	\]
By first decomposing the directional derivative into its radial and
angular parts, then partially integrating, and finally recombining
terms, we obtain the relation
	\begin{eqnarray}
	&&{1\over r} \int_{{\rm U}(1|1)} (D_0 f \circ \phi)(k,r) dk =
	\nonumber \\
	&&{\partial\over\partial r} \left( {1\over r} \int_{
	{\rm U}(1|1)} \langle k^{-1}\cdot X_0 , {\rm e}_1 \rangle
	(f\circ\phi)(r,k) dk \right) \nonumber
	\end{eqnarray}
where ${\rm e}_1$ is the unit vector ${\rm e}_1 = \pmatrix{1\cr
0\cr}$, and $k^{-1}\cdot X_0$ denotes the element $X_0$ rotated by
$k^{-1}$.  We now integrate the total radial derivative to get
	\[
	{\cal R}[D_0 f] = \lim_{r\to 0} {1\over r}
	\int_{{\rm U}(1|1)} \langle k^{-1}\cdot X_0 ,
	{\rm e}_1 \rangle (f\circ\phi)(k,r) dk .
	\]
By Taylor expanding $f$ to linear order in $r$ and doing the resulting
integral over ${\rm U}(1|1)$, we obtain the expression
	\[
	{\cal R}[D_0 f] = \langle X_0 , {\rm grad} f(0) \rangle
	= (D_0 f)(0),
	\]
from which conclude
	\[
	{\cal R}[f] = f(0) .
	\]
This elementary result is well-known; for a direct derivation see e.g.
\cite{Constantinescu}. The above construction has the great virtue of
using no more than the generic structures at hand. It therefore
generalizes easily to more complicated situations. In particular, it
can be used to construct the boundary terms associated with the
polar-coordinate integral (\ref{polar}). The result is as stated in
the text.

\end{multicols}
\end{document}